\documentclass[journal]{IEEEtran}

\usepackage{amsmath}
\usepackage{subfigure} 
\usepackage{lineno,hyperref}
\usepackage{xcolor}
\usepackage{extarrows}
\usepackage{algorithm}
\usepackage{algorithmic}
\usepackage{amssymb}
\usepackage{bm}
\usepackage{graphicx}
\usepackage[justification=centering]{caption}
\newcommand{\rev}{\textcolor{black}}
\newcommand{\rea}{\textcolor{black}}


\usepackage{amsthm}

\begin{document}
\title{Unitary Approximate Message Passing for Sparse Bayesian Learning} 
\author{Man~Luo, Qinghua~Guo, \IEEEmembership{Senior Member, IEEE}, Ming~Jin, Yonina C. Eldar, \IEEEmembership{Fellow, IEEE}, Defeng (David) Huang, \IEEEmembership{Senior Member, IEEE}, and Xiangming Meng   \\
\thanks{Part of this work was presented in the 16th IEEE APWCS 2019. 
Corresponding author: Qinghua Guo.}
\thanks{M. Luo and Q. Guo are with the School of Electrical Computer and Telecommunications Engineering, University of Wollongong, Australia, (e-mail: ml857@uowmail.edu.au, qguo@uow.edu.au).}
\thanks{M. Jin is with the Faculty of Electrical Engineering
and Computer Science, Ningbo University, Ningbo, China (e-mail: jinming@nbu.edu.cn).}
\thanks{Y. C. Eldar is with the Faculty of Math and CS, Weizmann Institute of Science, Rehovot, Israel (email: yonina.eldar@weizmann.ac.il).}
\thanks{D. Huang is with the School of Engineering, University of Western Australia, Perth Australia (e-mail: david.huang@uwa.edu.au).}
\thanks{X. Meng is with the Institute for Physics of Intelligence, The University of Tokyo, Hongo, Tokyo 113-0033, Japan (email: meng@g.ecc.u-tokyo.ac.jp). }
}
 
\markboth{Unitary Approximate Message Passing for Sparse Bayesian Learning}
{Shell \MakeLowercase{\textit{et al.}}: Bare Demo of IEEEtran.cls for IEEE Journals}

\maketitle

\begin{abstract}
Sparse Bayesian learning (SBL) can be implemented with low complexity based on the approximate message passing (AMP) algorithm. However, it \rev{does not work well for a generic measurement matrix}, which may cause AMP to diverge. Damped AMP has been used for SBL to alleviate the problem at the cost of reducing convergence speed. In this work, we propose a new SBL algorithm based on structured variational inference, leveraging AMP with a unitary transformation (UAMP). Both single measurement vector and multiple measurement vector problems are investigated. It is shown that, compared to state-of-the-art AMP-based SBL algorithms, the proposed UAMP-SBL is more robust and efficient, leading to remarkably better performance. 
  
\end{abstract}


\IEEEpeerreviewmaketitle

\section{Introduction}

 
We consider the problem of recovering a sparse signal $\mathbf{x}$ from noisy measurements $\mathbf{y} = \mathbf{Ax} + \mathbf{w}$, where $\mathbf{A}$ is a known measurement matrix \cite{book}. This problem finds numerous applications in various areas of signal processing, statistics and computer science. One approach to recovering $\mathbf{x}$ is to use sparse Bayesian learning (SBL), where $\mathbf{x}$ is assumed to have a sparsity-promoting prior \cite{tipping2001sparse}.       
Conventional implementation of SBL involves matrix inversion in each iteration, resulting in prohibitive computational complexity for large scale problems. 

The approximate message passing (AMP) algorithm \cite{Donoho}, \cite{donoho2009message} has been proposed for low-complexity implementation of SBL \cite{al2014sparse, Zhu2018AMP}. AMP was originally developed for compressive sensing based on loopy belief propagation (BP) \cite{donoho2009message}. Compared to convex optimization based algorithms such as LASSO \cite{tibshirani2011regression} and greedy algorithms such as iterative hard-thresholding \cite{blumensath2010normalized}, AMP has low complexity and its performance can be rigorously characterized by a scalar state evolution (SE) in the case of a large independent and identically distributed (i.i.d.) (sub-)Gaussian matrix $\mathbf{ A }$ \cite{javanmard2013state}. AMP was later extended in \cite{rangan2011generalized} to solve general estimation problems with a generalized linear observation model \cite{meng2018unified}. By implementing the E-step using AMP in the expectation maximization (EM) \rev{based} SBL method, matrix inversion can be avoided, leading to a significant reduction in computational complexity. However, AMP does not work well for a generic matrix such as non-zero mean, rank-deficient, correlated, or ill-conditioned matrix $\mathbf{A}$ \cite{rangan2019convergence}, resulting in divergence and poor performance.

Many variants to AMP have been proposed to address the divergence issue and achieve better robustness to \rev{a generic} $\mathbf{ A }$, such as the damped AMP \cite{rangan2019convergence}, swept AMP \cite{manoel2014sparse}, \rev{generalized approximate message passing algorithm (GAMP)} with adaptive damping \cite{vila2015adaptive},  vector AMP \cite{rangan2017vector}, orthogonal AMP \cite{ma2017orthogonal}, \rev{memory AMP \cite{MAMP}, convolutional AMP \cite{CAMP}} and more. In \cite{al2018gamp}, \rev{by incorporating damped Gaussian generalized AMP (GGAMP) to the EM-based SBL method}, a GGAMP-SBL algorithm was proposed. Although the robustness of the approach is significantly improved, it comes at the cost of slowing the convergence. \rev{In addition, the algorithm} still exhibits significant performance gap from the support-oracle bound when the measurement matrix has relatively high correlation, large condition number or non-zero mean.    

For a general linear inverse problem, \cite{guo2015approximate, BiUTAMP} proposed to apply AMP to a unitary transform of the original model, 
where \rev{the unitary matrix for the transformation can be} obtained by the singular value decomposition (SVD) of $\mathbf{A}$. In the case of a circulant $\mathbf{A}$, \rev{the normalized discrete Fourier transform matrix can be used for the unitary transformation}, enabling highly efficient implementation with the fast Fourier transform (FFT) algorithm \cite{guo2013}.
This leads to an AMP variant named unitary AMP (UAMP), which was formerly called  AMP with unitary transformation (UTAMP). \footnote{ \rea{SVD plays an important role in both UAMP and VAMP. In UAMP, SVD is used to obtain the unitary transformed model. In VAMP, the linear minimum mean squared error (LMMSE) estimator  results in cubic complexity in each iteration, and VAMP relies on SVD to implement the LMMSE estimator with low complexity.}} 
\rev{In this work}, we apply this concept to SBL, resulting in a new  SBL algorithm called UAMP-SBL. UAMP-SBL achieves more efficient sparse signal recovery with significantly enhanced robustness, compared to the state-of-the-art AMP-based SBL algorithm GGAMP-SBL \cite{al2018gamp}.

To develop UAMP-SBL, we apply structured variational inference (SVI) \cite{jordan1999introduction}, \cite{winn2005variational}, \cite{xing2012generalized}.
In particular, the formulated problem is represented by a factor graph model, based on which approximate inference is implemented in terms of structured variational message passing (SVMP) \cite{winn2005variational}, \cite{xing2012generalized}, \cite{dauwels2007variational}. The use of SVMP allows the incorporation of UAMP to the message passing algorithm to handle the most computational intensive part of message computations with high robustness and low complexity. 
In UAMP-SBL, a Gamma distribution is used as the hyperprior for the precisions of the elements of $\mathbf{x}$. \rev{We propose to tune the shape parameter of the Gamma distribution} automatically during iterations. 
We show by simulations that, in many cases \rev{ with a  generic measurement matrix}, UAMP-SBL can still approach the support-oracle bound closely. We also investigate \rev{SE-based performance prediction for UAMP-SBL and analyze the impact of the shape parameter on SBL.}
In addition, \rev{the UAMP-SBL algorithm is extended from single measurement vector} (SMV) problems to multiple measurement vector (MMV) problems \cite{jin2013support}, \cite{Eldarv}, \cite{Eldarv1}. 
Based on our preliminary results in {\cite{luo2019sparse}}\footnote{\rev{Compared to {\cite{luo2019sparse}}, we present a new derivation of UAMP-SBL, extend it from SMV to MMV, and provide theoretical analyses and comprehensive comparisons.}}, UAMP-SBL was applied to inverse synthetic aperture radar (ISAR) \cite{kang2020pattern}, where the measurement matrix can be highly correlated in order to achieve high Doppler resolution. Real data experiments in \cite{kang2020pattern} demonstrate its superiority in terms of both recovery performance and speed.

The rest of the paper is organized as follows. \rev{We briefly introduce SBL and (U)AMP in Section II. \rea{In Section III}, UAMP-SBL is derived for SMV problems and SE-based performance prediction for UAMP-SBL is also discussed. The impact of the shape parameter is analyzed in Section IV.} UAMP-SBL is extended to the MMV setting in \rev{Section V}. Numerical results are provided in \rev{Section VI}, followed by conclusions in \rev{Section VII}.

Throughput the paper, we use boldface lowercase and uppercase letters to represent column vectors and matrices, respectively. 
The superscript $(\cdot)^H$ represents the conjugate transpose for a complex matrix, and the transpose for a real matrix. 
We use $\mathbf{1}$ and $\mathbf{0}$ to denote an all-one vector and an all-zero vector with proper sizes, respectively. 
The notation $\mathcal{N} ( \mathbf{x}| \bm{\mu},\bm{\Sigma} )$ denotes a Gaussian distribution of $\mathbf{x}$ with mean $\bm{\mu}$ and covariance  $\bm{\Sigma}$, and ${\mathrm{Ga}(\bm{\gamma}| \epsilon,\eta)}$ is a Gamma distribution with shape parameter $\epsilon$ and rate parameter $\eta$. We use $ \left|\cdot \right|^{.2}$ to denote the element-wise magnitude squared operation, and $ \left\|\cdot \right\|$ for the $l_2$ norm. The notation $\left\langle f(\mathbf{x}) \right\rangle _{q(\mathbf{x})}$ denotes the expectation of $f(\mathbf{x})$ with respect to probability density function $q(\mathbf{x})$, and $E[ \cdot ]$ is the expectation over all random variables involved in the brackets. We use $Diag(\mathbf{a})$ to represent a diagonal matrix with elements of $\mathbf{a}$ on its diagonal, $Z_{m,n}$ is the $(m,n)$th element of $\mathbf{Z}$, and $a_n$ is the $n$th element of vector $\mathbf{a}$. The element-wise product and division of two vectors $\mathbf{a}$ and $\mathbf{b}$ are written as $\mathbf{a}\cdot\mathbf{b}$ and $\mathbf{a}./\mathbf{b}$, respectively. The superscript $\mathbf{a}^t$ is the $t$th iteration in an iterative algorithm.

\section {Background} 

\subsection{Sparse Bayesian Learning}
Consider recovering a length-$N$ sparse vector $\mathbf{x}$ from measurements  
\begin{equation}
 \mathbf{y} = \mathbf{Ax} + \mathbf{w}, \\
\label{y=ax+w}
\end{equation}
where $\mathbf{y}$ is a measurement vector of length $M$, the measurement matrix $\mathbf{A}$ has size $M \times N$, 
$\mathbf{w}$ denotes a Gaussian noise vector with mean zero and covariance  matrix $\beta^{-1}\mathbf{I}$, and $\beta$ is the precision of the noise. It is assumed that the elements in $\mathbf{x}$ are independent and the following two-layer sparsity-promoting prior is used 
\begin{eqnarray}
p(\mathbf{x}|\bm{\gamma})&=&\prod_{n}p(x_n|\gamma_n)=\prod_{n} \mathcal{N} ({x_n}| 0 , \gamma_n^{-1}), \label{abcdea} \\
p(\bm{\gamma})&=&\prod_{n}p(\gamma_n)=\prod_{n}  \mathrm{Ga}(\gamma_n|\epsilon,\eta), \label{abcdeb}
\end{eqnarray}
i.e., the prior of $x_n$ is a Gaussian mixture 
\begin{equation}
p(x_n)=\int \mathcal{N} ({x_n}| 0 , \gamma_n^{-1}) p(\gamma_n)d\gamma_n,
\end{equation}
\rea{where the precision vector $\mathbf{\gamma}=[\gamma_1, \gamma_2, ..., \gamma_N]^H$.}


\rea{In the conventional SBL algorithm by Tipping \cite{tipping2001sparse}, the precision vector $\bm \gamma$ is learned by maximizing the a posteriori probability 
 	\begin{equation}	
 	p(\bm \gamma|\mathbf{y}) \propto p(\mathbf{y}|\bm\gamma) p(\bm{\gamma}),    
 	\end{equation}
where the marginal likelihood function 
 	\begin{eqnarray}
 	p(\mathbf{y}|\bm{\gamma})=\int p(\mathbf{y}|\mathbf{x})  p(\mathbf{x}|\bm{\gamma}) d\mathbf{x}.
 	\end{eqnarray}
It can be shown that \cite{tipping2001sparse} 
\begin{eqnarray}
\log{p(\mathbf{y}|\bm{\gamma})} 
\!\!\!\!\!\!&=&\!\!\!\!\!\! -\frac{1}{2} \left(\log{ | \mathbf{B}|}  + 
\mathbf{y}^H \mathbf{B}^{-1} \mathbf{y}\right) + const \\
&=& \frac{1}{2} \big( \log{ |\mathbf{\Sigma}|} +\log{ |Diag(\bm{\gamma})|} \nonumber \\
&&~~~~~~~~~~~~~~~- \bm{\zeta}^H Diag(\bm{\gamma})\bm{\zeta} \big) +const1, \label{employed}
\end{eqnarray}
where $const$ and $const1$ represent terms independent of $\bm{\gamma}$, and 
\begin{eqnarray}
\mathbf{B} &=& \beta^{-1} \mathbf{I} + \mathbf{A} Diag(\bm{\gamma})^{-1} \mathbf{A}^H,  \\ 
\mathbf{\Sigma}&=&\left(\beta \mathbf{A}^H\mathbf{A}+Diag({\bm{\gamma}})\right)^{-1}, \label{employa} \\
\bm{\zeta}&=&\beta \mathbf{\Sigma A}^H\mathbf{y} \label{employb}.
\end{eqnarray} 
The a posteriori probability of $\mathbf{x}$  
\begin{equation}
p(\mathbf{x}|\mathbf{y},\bm{\gamma})=\mathcal{N}(\mathbf{x}|\bm{\zeta},\bm{\Sigma}).
\end{equation}
By taking the logarithm of $p(\bm\gamma|\mathbf{y})$ and ignoring terms independent of $\bm\gamma$, the learning of $\bm{\gamma}$ is to maximize the following objective function \cite{tipping2001sparse} 
 	\begin{eqnarray}
 	\mathcal{L}(\bm{\gamma})= \log{p(\mathbf{y}|\bm{\gamma})} +  \sum_{n=1}^N (  \epsilon \log{\gamma_n} - \eta \gamma_n).
 	\end{eqnarray}
As the value of $\bm{\gamma}$ that maximizes $\mathcal{L}(\bm{\gamma})$ cannot be obtained in a closed form, iterative re-estimation is employed by taking advantage of \eqref{employed},
i.e., with a learned ${\bm{\gamma}}$ in the last iteration, 
compute $\bm{\Sigma}$ and $\bm{\zeta}$ with \eqref{employa} and \eqref{employb}, then update $\bm{\gamma}$ by maximizing $\mathcal{L}(\bm{\gamma})$ with \eqref{employed} used, which leads to a closed form to update $\gamma_n$
\begin{equation}
\gamma_n=(2 {\epsilon} +1)/(2 \eta+|\zeta_n|^2 + \Sigma_{n,n}), n=1,..., N.
\end{equation}
In summary, Tipping's SBL algorithm} (which is called SBL hereafter) executes the following iteration \cite{tipping2001sparse}:    
\begin{eqnarray}
&&\!\!\!\!\!\!Repeat \nonumber \\
&&~\mathbf{Z}=\left(\beta \mathbf{A}^H\mathbf{A}+Diag(\hat{\bm{\gamma}})\right)^{-1} \label{adda} \\
&&~\mathbf{\hat x}=\beta \mathbf{ZA}^H\mathbf{y} \\
&&~\hat\gamma_n=(2 {\epsilon} +1)/(2 \eta+|\hat x_n|^2 + Z_{n,n}), n=1,..., N.  \\
&&\!\!\!\!\!\!Until~terminated \nonumber 
\end{eqnarray}
\rea{If the noise precision $\beta$ is unknown, its estimation can be incorporated as well. The SBL algorithm can also be derived based on the EM algorithm \cite{tipping2001sparse}, \cite{al2018gamp}}. The SBL algorithm requires a matrix inverse in \eqref{adda} in each iteration. This results in cubic complexity in each iteration, which can be prohibitive for large scale problems. To address this issue, the implementation of the E-step using AMP has been investigated. GGAMP was used in \cite{al2018gamp} to implement the E-Step, where sufficient damping is used to enhance the robustness of the algorithm against \rev{a generic measurement matrix}. This leads to the GGAMP-SBL algorithm \rev{with complexity significantly lower than that of SBL.} 

\subsection{(U)AMP}

AMP was derived based on the loopy BP with Gaussian and Taylor-series approximations \cite{donoho2009message}, {\cite{rangan2011generalized}}, which can be used to efficiently solve linear inverse problems due to its low complexity.
An issue with AMP is that it can easily diverge in the case of \rea{a generic measurement matrix, such as} correlated, ill-conditioned, non-zero mean or rank-deficient $\mathbf{A}$ \cite{rangan2019convergence}.  
Inspired by the work in \cite{guo2013}, it was shown in \cite{guo2015approximate} that the robustness of AMP is remarkably improved through simple pre-processing, i.e., performing a unitary transformation to the original linear model \cite{guo2015approximate}, \cite{BiUTAMP}. 
As any matrix $\mathbf{A}$ has an SVD  $\mathbf{A= U \Lambda V}$ with $\mathbf{U}$ and $\mathbf{V}$ being two unitary matrices, performing a unitary transformation with $\mathbf{U}^H$ leads to the following model
\begin{equation}
\mathbf{r=\Phi x}+\bm{\omega},
\label{r=uy}
\end{equation} 
where $\mathbf{r=U}^H \mathbf{y}$, $\mathbf{\Phi}=\mathbf{U}^H\mathbf{A}=\mathbf{\Lambda}\mathbf{V}$,
$\mathbf{\Lambda}$ is an $M\times N$ rectangular diagonal matrix, and $\bm{\omega} = \mathbf{U}^H \mathbf{w}$ remains a zero-mean  Gaussian noise vector with the same covariance matrix  $\beta^{-1} \mathbf{I}$. 
Applying the vector step size AMP {\cite{rangan2011generalized}} with model \eqref{r=uy} leads to the first version of UAMP \rev{(called UAMPv1)} shown in Algorithm \ref{UTAMPv1}.\footnote{\rev{By replacing $\mathbf{r}$ and $\mathbf{\Phi}$ 
with $\mathbf{y}$ and $\mathbf{A}$ in Algorithm \ref{UTAMPv1} respectively, the original AMP algorithm is recovered.}}
\rev{    
\begin{algorithm}
	\caption{UAMP (UAMPv2 executes operations in [ ])}
	Initialize $\bm{\tau}_x^{(0)} (\mathrm{or}~{\tau}_x^{(0)})>0$ and ${{\mathbf{x}}^{(0)} }$. Set $\mathbf{s}^{(-1)}=\mathbf{ 0 }$ and $t=0$. Define vector $\bm{\lambda}=\mathbf{ \Lambda \Lambda}^H \textbf{1}$.\\
	\textbf{Repeat}   
	\begin{algorithmic}[1]
		\STATE $\bm{\tau}_p$ = $   \mathbf{|\Phi|}^{.2} \bm{\tau}^t_x$~~~~~~~~~~~~~ $\left[\mathrm{or}~ \bm{\tau}_p =  \tau^t_x  \bm{\lambda}\right]$ \\
		\STATE $ \mathbf{p}= \mathbf{\Phi}  {{\mathbf{x}}^{t} } - \bm{\tau}_{p} \cdot  \mathbf{s}^{t-1} $\\
		\STATE $ \bm{\tau}_s = \mathbf{1}./ (\bm{\tau}_p+\beta^{-1} \mathbf{1}) $\\
		\STATE $ \mathbf{s}^t= \bm{\tau}_s \cdot (\mathbf{r}-\mathbf{p}) $\\
		\STATE $\mathbf{1}./ \bm{\tau}_q$ = $ |\mathbf{\Phi}^H |^{.2}  \bm{\tau}_s$~~~~~~~~$\left[\mathrm{or}~ \mathbf{1}./ \bm{\tau}_q = (\frac{1}{N} \bm{\lambda}^H \bm{\tau }_s) \mathbf{1}\right]$ \\
		\STATE $ \mathbf{q} = {{\mathbf{x}}^{t} } + \bm{\tau}_q \cdot(  \mathbf{\Phi}^H \mathbf{s}^t)$\\
		\STATE $\bm{\tau}_x^{t+1}$ = $\bm{\tau}_q \cdot g_{x}' ( \mathbf{q}, \bm{\tau}_q)$~~~~~$\left[\mathrm{or}~ \tau_x^{t+1} \!=\!  \frac{1}{N}  \mathbf{1}^H  \left(\bm{\tau}_q \cdot g_{x}' ( \mathbf{q}, \tau_q)\right) \right]$ \\
		\STATE 	$ {{\mathbf{x}}^{t+1} } = g_{x}  ( \mathbf{q}, \bm{\tau}_q)$\\	
		\STATE 	$  t=t+1$
	\end{algorithmic}
	\textbf{Until terminated}   
	\label{UTAMPv1}
\end{algorithm}}

We can apply an average operation to two vectors: $\bm{\tau}_x$ in Line 7 and $|\mathbf{\Phi}^H |^{.2}  \bm{\tau}_s$ in Line 5 of \rev{UAMPv1 in} Algorithm \ref{UTAMPv1}, leading to the second version of UAMP \cite{guo2015approximate} \rev{(called UAMPv2), where the operations in the brackets of Lines 1, 5 and 7 are executed} (refer to \cite{BiUTAMP} for the derivation). 
Compared to AMP and \rev{UAMPv1,  UAMPv2 does not require matrix-vector products in Lines 1 and 5,} so that the number of matrix-vector products is reduced from 4 to 2 per iteration. This is a significant reduction in computational complexity because the complexity of AMP-like algorithms is dominated by matrix-vector products. 

In the (U)AMP algorithms, $g_x(\mathbf{q}, \bm{\tau}_q )$ is related to the prior of $\mathbf{x}$ and returns a column vector with the $n$th
element $[ g_x(\mathbf{q}, \bm{\tau}_q ) ]_n$ given by 
\begin{equation}
[g_x(\mathbf{q}, \bm{\tau}_q ) ]_n
=
\frac{\int x_n p(x_n) \mathcal{N} (x_n ; q_n, \tau_{q_n})  d x_n }{\int  p(x_n) \mathcal{N} (x_n ; q_n, \tau_{q_n})  d x_n },
\label{g_x}
\end{equation}
where we note that $p(x_n)$ represents a general known prior for $x_n$.
The function $g_x'(\mathbf{q},\bm{\tau}_q)$ returns a column vector and the $n$th element is denoted by $[ g_x'(\mathbf{q}, \bm{\tau}_q ) ]_n$, where the derivative is taken with respect to $q_n$. 


%

\section{Sparse Bayesian Learning Using UAMP}

\subsection{Problem Formulation and Approximate Inference}
 
To enable the use of UAMP, we employ the unitary transformed model $\mathbf{r=\Phi x}+\bm{\omega}$ in \eqref{r=uy}. As in many applications the noise precision $\beta$ is unknown, its estimation is also considered. The joint conditional distribution of ${\mathbf{x}}$, $\bm{\gamma}$ and $\beta$ can be expressed as
\begin{eqnarray}
p(\mathbf{x},\bm{\gamma}, \beta| \mathbf{r}) 
\propto p(\mathbf{r}|\mathbf{x},\beta)p(\mathbf{x}|\bm{\gamma})p(\bm{\gamma})p(\beta),
\end{eqnarray}    
where $p(\mathbf{x}|\bm{\gamma})$ and $p(\bm{\gamma})$ are given by \eqref{abcdea} and \eqref{abcdeb}, respectively. 
We assume an improper prior $p(\beta) \propto 1/\beta$  for the noise precision \cite{tipping2001sparse}. According to the \rev{transformed model \eqref{r=uy}}, $p(\mathbf{r}|\mathbf{x},\beta)=\mathcal{N} (\mathbf{r}| \mathbf{\Phi x}, \beta^{-1}\mathbf{I})$. Our aim is to find the marginal distribution $p(\mathbf{x}| \mathbf{r})$. The a posteriori mean is then used as an estimate of $\mathbf{x}$ in the sense of minimum mean squared error (MSE). However, exact inference is intractable due to the high dimensional integration involved, \rev{so we resort to approximate inference techniques.} 

\rev{Variational inference is a machine learning method for approximate inference, and it has been widely used to approximate posterior densities for Bayesian models \cite{jordan1999introduction}, \cite{winn2005variational}, \cite{xing2012generalized}. In variational inference, a trial density function is chosen and optimized by minimizing the Kullback-Leibler
(KL) divergence between the trial function and the true a posteriori function. Instead of using fully factorized trial functions where all variables are assumed to be independent (thereby likely resulting in poor approximations), more structured factorizations can be used, leading to SVI algorithms. With graphical models, SVI \rea{can be} formulated as message-passing algorithms \cite{winn2005variational}, \cite{xing2012generalized}, \cite{dauwels2007variational}, which is termed SVMP. In this work, SVMP is adopted because the use of SVMP facilitates the incorporation of the message passing algorithm UAMP into \rea{SVMP. We} will show how UAMP can be used to handle the most computational intensive part of message computations, \rea{enabling} us to achieve low complexity and high robustness. With SVMP, we can find an approximation to the marginal distribution $p(\mathbf{x}| \mathbf{r})$, where an approximation to $p(\bm{\gamma}| \mathbf{r})$ is also involved (the approximate inference for $\mathbf{x}$ and $\bm{\gamma}$ is performed alternately).}



\begin{figure}
	\centering
	\includegraphics[width=0.9\linewidth]{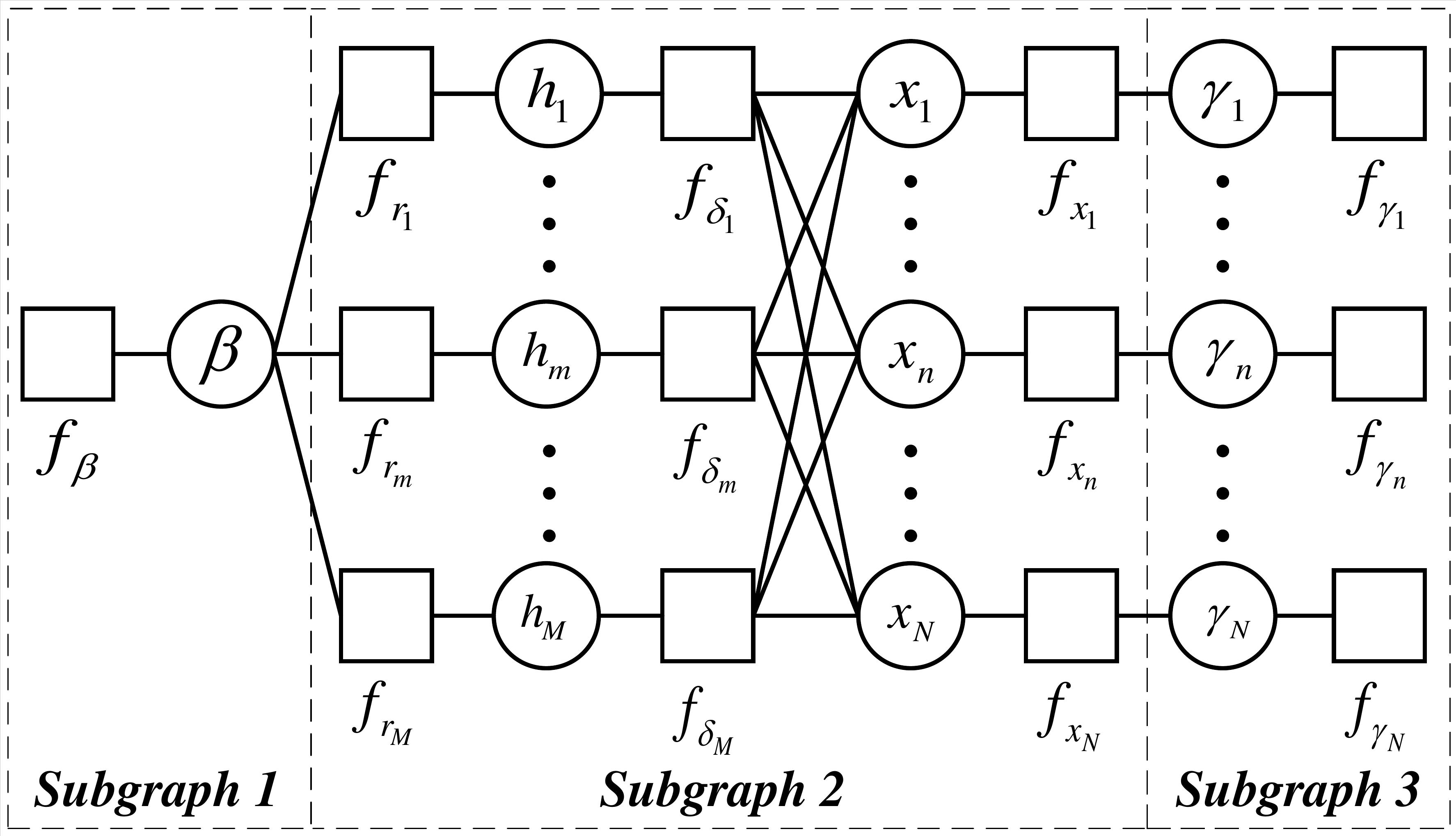}
	\centering
	\caption{Factor graph of (\ref{eq:factor1}) for deriving UAMP-SBL.}
	\label{fig:factor graph}
\end{figure}

We introduce an auxiliary variable $\mathbf{h}=\mathbf{\Phi x}$ 
to facilitate the incorporation of UAMP, which is crucial to an efficient realization of SBL. Then the conditional joint distribution is
\begin{eqnarray}
&&\!\!\!\!\!\!\!\!\!\!\!\!\!\!\!\!p(\mathbf{x},\mathbf{h},\bm{\gamma}, \beta| \mathbf{r}) \nonumber \\
&&\!\!\!\!\!\propto p(\mathbf{r}|\mathbf{h},\beta)p(\mathbf{h}|\mathbf{x})p(\mathbf{x}|\bm{\gamma})p(\bm{\gamma}|{\epsilon})p(\beta) \nonumber \\
&&\!\!\!\!\!= \prod_{m=1}^{M} \mathcal{N} (r_m | h_m, {\beta^{-1}}) \prod_{m=1}^{M} \delta(h_m-[\mathbf{\Phi}]_m  {\mathbf{x}}) \nonumber \\
&&~\prod_{n=1}^{N} \mathcal{N} (x_n | 0, \gamma_n^{-1})\prod_{n=1}^{N}  \mathrm{Ga}(\gamma_n|\epsilon,\eta) p(\beta).
\label{eq:factor1}
\end{eqnarray}
\rev{To facilitate the derivation of the message passing algorithm,
a factor graph} representation of the factorization in \eqref{eq:factor1} is shown in Fig.\ref{fig:factor graph}, \rev{where the local functions} $f_{\beta}(\beta) \propto 1/\beta$, $f_{r_m}({r_m}, h_m, \beta ) =\mathcal{N} (r_m | h_m, \beta^{-1})$, $ f_{\delta_m}(h_m, { \mathbf{x}} )  =  \delta(h_m-[\mathbf{\Phi}]_m {\mathbf{x}} ) $, 
$f_{x_n}({x_n}, \gamma_n)=\mathcal{N} (x_n |  0, \gamma_n^{-1})$, $f_{\gamma_n}(\gamma_n) =\mathrm{Ga}(\gamma_n|\epsilon,\eta) $ and $[\mathbf{\Phi}]_m$ is the $m$th row of matrix $\mathbf{\Phi}$.

Following SVI, we define the following structured trial function
\begin{equation}
\tilde q(\mathbf{x},\mathbf{h},\bm{\gamma}, \beta)= \tilde q(\beta) \tilde q(\mathbf{x},\mathbf{h}) \tilde q(\bm{\gamma}).
\end{equation} 
In terms of SVMP, the use of the above trial function corresponds to a partition of the factor graph shown by the dotted boxes in Fig. \ref{fig:factor graph}, where $\tilde q(\beta)$, $\tilde q(\mathbf{x},\mathbf{h})$ and $\tilde q(\bm{\gamma})$ are associated with Subgraphs 1, 2 and 3, respectively.
As the KL divergence
\begin{eqnarray}
\mathcal{KL}\big(\tilde{q}(\beta) \tilde{q}( \mathbf{x}, \mathbf{h}) \tilde{q}(\bm{\gamma}) || p(\mathbf{x},\mathbf{h},\bm{\gamma}, \beta| \mathbf{r})\big), 
\end{eqnarray} 
is minimized, it is expected that
\begin{eqnarray}
 \tilde{q}( \mathbf{x}, \mathbf{h}) &\approx&  p(\mathbf{x},\mathbf{h}| \mathbf{r}), \label{svi} \\
 \tilde{q}( \bm{\gamma}) &\approx&  p(\bm{\gamma}| \mathbf{r}),\\
 \tilde{q}( \beta) &\approx&  p(\beta| \mathbf{r}). 
\end{eqnarray} 
Integrating out $\mathbf{h}$ in \eqref{svi}, which corresponds to running BP in Subgraph 2 (except the factor nodes connecting external variable nodes), we have $\tilde{q}( \mathbf{x}) \approx p(\mathbf{x}| \mathbf{r})$. Running BP in Subgraph 2 involves the most intensive computations; fortunately it can be handled efficiently and with high robustness using UAMP. \rev{The derivation of UAMP-SBL is shown in Appendix A, \rea{and the algorithm} is summarized in Algorithm \ref{vector_UTAMP_SBL_SMV-table}. }      



\begin{algorithm}
	\caption{UAMP-SBL}
	Unitary transform: $\mathbf{r=U}^H \mathbf{ y }=\mathbf{\Phi x} +\bm{\omega}$, where $\mathbf{\Phi=U}^H\mathbf{A}=\mathbf{\Lambda V}$, and $\mathbf{A}$ has SVD $\mathbf{A=U \Lambda V}$.\\
	Define vector $\bm{\lambda}= \mathbf{\Lambda \Lambda}^H \mathbf{1}$.\\  
	Initialization: ${\tau_{x}^{(0)}=1}$, $\hat{\mathbf{x}}^{(0)}=\textbf{0}$, ${ {\epsilon}=0.001}$, $\hat{{\bm{\gamma}}}=\textbf{1}$, $\hat{ \beta}=1$, $\mathbf{s}=\mathbf{ 0 }$, and $t=0$.
	
	\textbf{Do}     
	\begin{algorithmic}[1]
		\STATE $ \bm{\tau}_p$ = $ { \tau^t_x }  \bm{\lambda}$\\
		\STATE $ \mathbf{p}= \mathbf{\Phi \hat x}^t - \bm{\tau}_{p} \cdot  \mathbf{s} $\\
		\STATE $ \mathbf{v}_h = \bm{\tau}_p./ (\bm{1}+\hat{ \beta}\bm{\tau}_p)$\\ 
		\STATE $\mathbf{\hat{h}}= (\hat\beta\bm{\tau}_p\cdot \mathbf{r}+\mathbf{p})./(\bm{1}+\hat\beta\bm{\tau}_p)$\\
		\STATE  $\hat{ \beta} =   {M}/   ( {  ||  \mathbf{r}-  \mathbf{\hat{h}}  ||^2   + \mathbf{1}^H\mathbf{v}_h  }  ) $\\	
		\STATE $ \bm{\tau}_s = \mathbf{1}./ (\bm{\tau}_p+\hat\beta^{-1} \mathbf{1}) $\\
		\STATE $ \mathbf{s}= \bm{\tau}_s \cdot (\mathbf{r}-\mathbf{p}) $\\
		
		\STATE 	$ 1/\tau_q = ({1}/{N}) \bm{\lambda}^H \bm{\tau }_s   $\\
		
		\STATE $ \mathbf{q} =  \mathbf{\hat x}^t + \tau_q \mathbf{\Phi}^H \mathbf{s}$\\
		\STATE 	$ {  \tau^{t+1}_x }=  (\tau_q/N) \mathbf{1}^H(\mathbf{1}. /(\mathbf{1}+\tau_q  \hat{{\bm{\gamma}}} ))$\\
		\STATE 	$   \hat{\mathbf{x}}^{t+1} = \mathbf{q}   ./(\mathbf{1}+\tau_q \hat{{\bm{\gamma}}} )$\\
		\STATE $   \hat{{\gamma}}_n  = ( {2 {\epsilon} +1})/( |{{\hat{x}}_n^{t+1} }|^2 +{\tau^{t+1}_x }), n=1, ...,N.$ \\
		\STATE ${\epsilon}=\frac{1}{2}\sqrt{\log(\frac{1}{N}\sum_{n}{\hat{\gamma}_n})-\frac{1}{N}\sum_{n}{\log{\hat{\gamma}}_n}}$  
		\STATE $t=t+1$ 
	\end{algorithmic}
	\textbf{while}	({$   || \hat{\mathbf{x}}^{t+1} - \hat{\mathbf{x}}^{t} ||^2/ ||  \hat{\mathbf{x}}^{t+1}  ||^2  > \delta_x$} and $t<t_{max}$)
	\label{vector_UTAMP_SBL_SMV-table}
\end{algorithm}

\rev{Regarding the UAMP-SBL \rea{in Algorithm \ref{vector_UTAMP_SBL_SMV-table}}, we have the following remarks:
\begin{enumerate}
\item[1.] UAMPv2 is employed in Algorithm 2. Similarly, UAMPv1 can also be used. By comparing UAMPv1 and UAMPv2, the differences lie in Lines 1, 8, 9 and 10 as vectors $\bm{\tau}^t_x$ and $\bm{\tau}_q$ need to be used. \rea{The UAMP-SBL algorithms with two version of UAMP} deliver comparable performance, but UAMP-SBL with UAMPv2 has lower complexity.        
\item[2.] In SBL with Gamma hyperprior, the shape parameter  $\epsilon$ and the rate parameter $\eta$ are normally chosen to be very small values {\cite{tipping2001sparse}}, and sometimes the value of the shape parameter $\epsilon$ is chosen empirically, e.g., $\epsilon=1$ in \cite{pedersen2012application}. In UAMP-SBL, we propose to tune the shape parameter automatically (as shown in Line 13) with the following empirical rule 
\begin{equation}
{\epsilon} =
\frac{1}{2}\sqrt{\log(\frac{1}{N}\sum_{n}{\hat{\gamma}_n})-\frac{1}{N}\sum_{n}{\log{\hat{\gamma}}_n}},
\label{eq_eps_update}
\end{equation}
i.e., $\epsilon$ is learned iteratively \rea{with the iteration}, starting from a small \rea{positive} initial value. 
We note that, as the log function is concave, the parameter $\epsilon$ in \eqref{eq_eps_update} is guaranteed to be non-negative.  
In Section IV, we will show that the shape parameter $\epsilon$ in the SBL algorithms functions as a selective amplifier for $\{\gamma_n\}$, and a proper $\epsilon$ plays a significant role in promoting sparsity, leading to considerable performance improvement. 
\end{enumerate}
}

\begin{figure}[t!]
	\centering
	\includegraphics[width=1\linewidth]{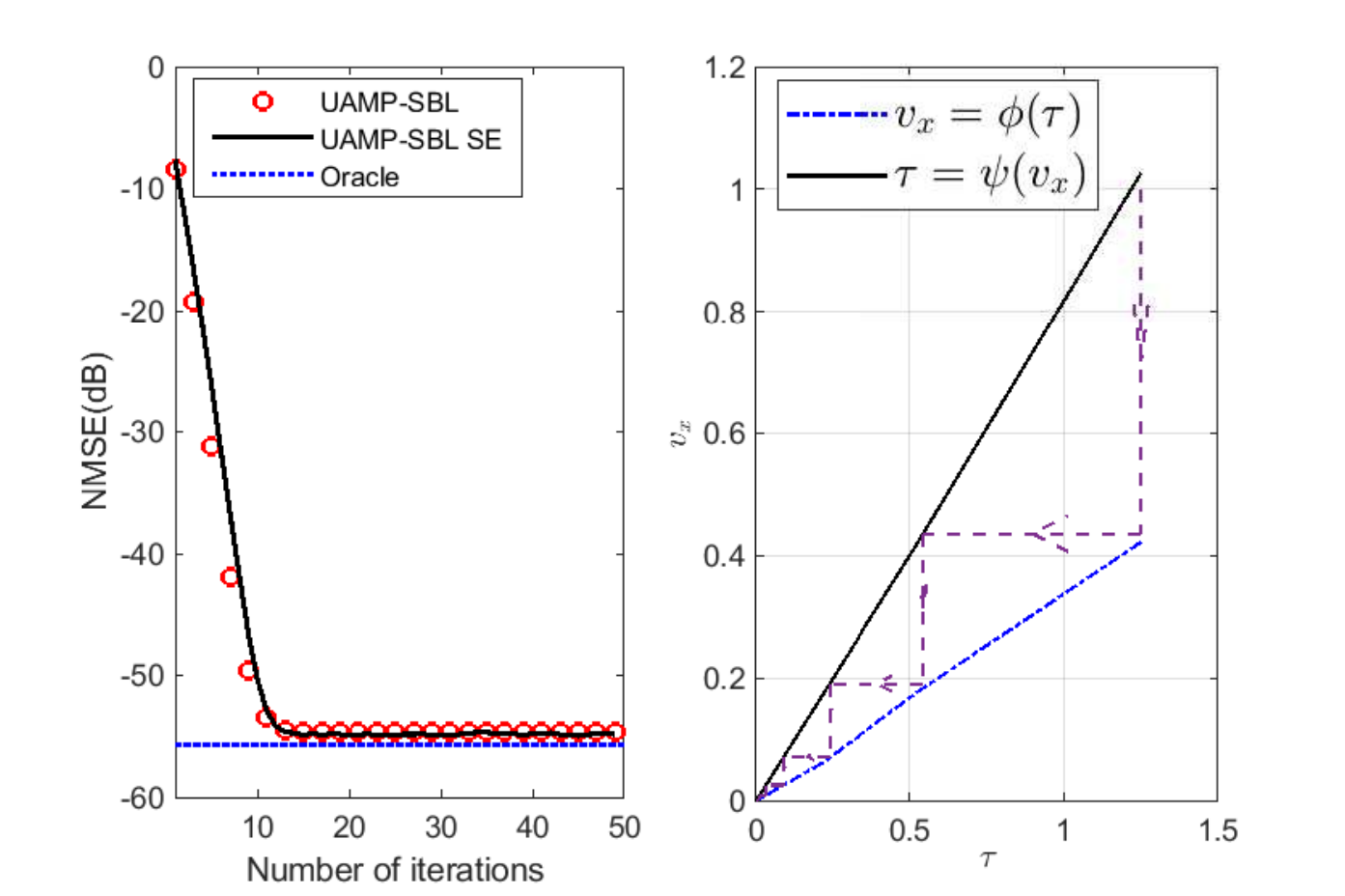}
	\caption{SE and evolution trajectory of UAMP-SBL with a nonzero mean $\textbf{A}$ ($N = 10240$, $M = 8192$, SNR = 50dB, sparsity rate $\rho=0.1$ and matrix mean $\mu = 10$).}
	\label{UAMPSBL_SE}
\end{figure}
 
\subsection{\rev{SE-Based Performance Prediction}}

In this section, leveraging UAMP SE, \rev{we study how to predict the performance of UAMP-SBL empirically.} 
We treat UAMP-SBL as UAMP with a special denoiser, enabling the use of UAMP SE to predict the performance of UAMP-SBL. The denoiser in the UAMP-SBL corresponds to Lines 10-13 of the UAMP-SBL algorithm (Algorithm \ref{vector_UTAMP_SBL_SMV-table}).

{As (U)AMP decouples the estimation of vector $\mathbf{x}$, in the $t$th iteration, we have the following pseudo observation model
	\begin{equation}
	q_n^t=x_n+w^t_n,
	\label{awgn}
	\end{equation}
	where $q_n^t$ is the $n$th element of $\mathbf{q}$ in Line 9 of the UAMP-SBL algorithm in the $t$th iteration, and $w^t_n$ denotes a Gaussian noise with mean 0 and variance $\tau^t$, which is given as  
	\begin{eqnarray}
	\tau^t = \frac{N}{\bm{1}^H \big(\bm{\lambda}./(v_x^{t}\bm{\lambda}+\beta^{-1}\bm{1})\big)}. 
	\end{eqnarray} 
	Here $v^t_x$ is the average MSE of $\{x_n\}$ after denoising in the $t$th iteration. As it is difficult to obtain a closed form for the average MSE, 
	we simulate the denoiser with the additive Gaussian \rea{noise} model \eqref{awgn} by varying the variance of noise $\tau^t$ (or the SNR), so that we can get a ``function" in terms of a table, with the variance of the noise as the input and the MSE as the output, i.e.,   
\begin{equation}
v_x=\phi(\tau).
\end{equation}  
The function $\phi(\cdot)$ is independent of  the measurement matrix $\mathbf{A}$. \rev{The performance of UAMP-SBL} can be predicted using the following iteration with the initialization of $v_x$:} 
\begin{equation}
\begin{aligned}
&Repeat \\
&~~~~~~~~ \tau = \frac{N}{\bm{1}^H \big(\bm{\lambda}./(v_x\bm{\lambda}+\beta^{-1}\bm{1})\big)} \triangleq \psi(v_x) \\
&~~~~~~~~v_x=\phi(\tau)  \\
&Until~terminated
\end{aligned}
\end{equation}
We show the predicted performance, simulated performance \rev{in terms of normalized MSE (NMSE, which is defined in \eqref{NMSE})} and the evolution trajectory of UAMP-SBL in Fig. \ref{UAMPSBL_SE} for a non-zero mean \rev{ measurement matrix} $\mathbf{A}$. It can be seen that the predicted performance matches well the simulated performance.

\subsection{\rev{Computational Complexity}}

UAMP-SBL works well with a simple single loop iteration, which is in contrast to the double loop iterative algorithm GGAMP-SBL \cite{al2018gamp}. The complexity of UAMP-SBL \rev{(with UAMPv2)} is dominated by two matrix-vector product operations in Line 2 and Line 9, i.e., $\mathcal{O} (MN)$ per iteration. The algorithm typically converges fast and delivers outstanding performance as shown in \rev{Section VI}. \rea{UAMP-SBL involves} an SVD , but it only needs to be computed once and may be carried out off-line. The complexity of economic SVD is $\mathcal{O} (\min\{M^2N, MN^2\})$. Note that for the runtime comparison in \rev{Section VI, we do not assume off-line SVD computation, and the time consumed by SVD is counted for UAMP-SBL.}

\section{Impact of the Shape Parameter $\epsilon$ in SBL}

In this section, we analyze the impact of the hyperparameter $\epsilon$ on the convergence of SBL. 
We focus on the case of an identity matrix $\mathbf{A}$. 
The same results for a general $\mathbf{A}$ are demonstrated numerically.	    

We consider the conventional SBL algorithm ($\eta$ is set to be zero) \cite{tipping2001sparse}. In the case of identity matrix $\mathbf{A}$, it reduces to 
\begin{eqnarray}
&&\!\!\!\!\!\!Repeat \nonumber \\
&&~~~~~Z_{n,n}=\left(\beta+{\gamma}_n^t \right)^{-1} \nonumber  \\
&&~~~~~\hat x_n=\beta Z_{n,n}y_{n}  \\
&&~~~~~\gamma_n^{t+1}=(2{\epsilon} +1)/(|\hat x_n|^2 + Z_{n,n}) \nonumber \\
&&\!\!\!\!\!\!Until~terminated \nonumber
\end{eqnarray}
\rev{Here note that in the above iteration we initialize $\gamma_n^{(0)}>0$.} The iteration in terms of $\gamma_n$ has a closed form
\begin{align}
\gamma^{t+1}_n & = \frac{2\epsilon +1}{(\beta (\beta + \gamma^t_n  )^{-1}y_n )^2 +(\beta + \gamma^t_n  )^{-1}} \nonumber  \\
& =(2\epsilon +1) \frac{ (\beta + \gamma^t_n  )^{2} }{(\beta y_n )^2 +\beta + \gamma^t_n } \label{iteration} \\
&\triangleq g_{\epsilon}(\gamma^t_n). \nonumber
\end{align}
 Next, we investigate the impact of $\epsilon$ on \rev{the convergence behavior and fixed point of the iteration \eqref{iteration} when $\epsilon=0$ or $\epsilon$ takes a positive value.}
 	
\rev{For the iteration \eqref{iteration} with a small positive initial value $\gamma_n^{(0)}$, we have the following proposition and theorem.}

\textbf{\textit{Proposition 1}}: \rev{When $ \epsilon = 0 $,  if $\beta y^2_n > 1$,} $\gamma_n^t$ converges to a stable fixed point 
\begin{equation}
\gamma'_{n} =  \frac{\beta}{\beta y^2_n - 1 }; \label{pa}
\end{equation}
if $\beta y^2_n  \leq 1$, $\gamma_n^t$  goes to $+\infty$. 
\begin{proof}  
	See Appendix B.
\end{proof}

\textbf{\textit{Theorem 1}}: \rev{ When $ \epsilon > 0$, {if $\beta y^2_n > 1+ 4\epsilon +4\sqrt{\epsilon^2 + \epsilon/2} $},}  $\gamma_n^t$ converges to a stable fixed point 
\begin{equation}
\gamma_{n(a)} =  \frac{ 2\beta(1+2\epsilon)}{\beta y_n^2 -4\epsilon - 1 + \sqrt{ \beta^2 y_n^4 - 8\epsilon \beta y_n^2 - 2\beta y_n^2 +1  } }; \label{pb}
\end{equation}
if $\beta y^2_n < 1+ 4\epsilon +4\sqrt{\epsilon^2 + \epsilon/2} $, $\gamma_n^t$  goes to $+\infty$.

\begin{proof}  
	See Appendix C.
\end{proof}

\rev{Based on Proposition 1 and Theorem 1, we make the following remarks:} 
\begin{enumerate}
\item[1.] \rev{If $\beta y^2_n  \leq 1$,  for both $ \epsilon = 0 $ and $ \epsilon > 0 $, $\gamma^t_n$ goes to $+\infty$. However, a positive $ \epsilon$ accelerates the move of $\gamma^t_n$ towards $+\infty$. This can be shown as follows. As $\beta > 0 $ and $\beta y^2_n  \leq 1$, we have $(\beta y_n)^2  \leq \beta$. Hence, from \eqref{iteration}}  
	\begin{align}
	\gamma^{t+1}_n =  g_{\epsilon}(\gamma^t_n)	&\geq (2 \epsilon +1)   \frac{ (\beta + \gamma^t_n  )^{2} }{ 2\beta + \gamma^t_n } \nonumber  \\ 
	& = (2 \epsilon +1) \left( \gamma^t_n  + \frac{ \beta^{2} }{ 2\beta + \gamma^t_n }\right) \nonumber \\
	&> (2 \epsilon +1)\gamma^t_n. \label{converge}   
	\end{align}
\rev{From \eqref{converge}, compared to $ \epsilon = 0 $, a positive value of $\epsilon$ \rea{moves $\gamma^t_n$ towards infinity more quickly.}   
Considering a fixed number of iterations, a positive value of $ \epsilon$ can be significant because the precision can reach a large value \rea{much faster}.
} 
\item[2.] \rev{When $ \epsilon = 0 $, $\gamma^t_n$ converges to a finite fixed point if $\beta y^2_n>1$. In contrast, when $ \epsilon> 0$, $\gamma^t_n$ goes to $+\infty$ \rea{if} $\beta y^2_n \in (1,  1+ 4\epsilon +4\sqrt{\epsilon^2 + \epsilon/2})$. \rea{This is an additional range for $\gamma^t_n$ to go to infinity}.  Hence, a positive $\epsilon$ is stronger in terms of promoting sparsity, \rea{compared to $\epsilon=0$.}}
\item[3.] \rev{ When $ \epsilon > 0 $, if $\beta y^2_n = 1+ 4\epsilon +4\sqrt{\epsilon^2 + \epsilon/2}$, $\gamma^t_n$ may converge or diverge because the iteration has a unique neutral fixed point as shown in Theorem 1.}
\item[4.] \rev{When $\beta y^2_n > 1+ 4\epsilon +4\sqrt{\epsilon^2 + \epsilon/2}$, $\gamma^t_n$ always converges to a fixed point. \rea{Based on \eqref{pa} and \eqref{pb}}, the ratio of the precisions obtained with $ \epsilon > 0 $ and $ \epsilon = 0 $ is given by}
\begin{equation}
\frac{\gamma_{n(a)}}{\gamma'_{n} } 
=
\frac{2(1+2\epsilon)}{1-\frac{4\epsilon}{\beta y^2_n -1} + \sqrt{\left(1-\frac{4\epsilon}{\beta y^2_n-1}\right)^2 - \frac{8\epsilon(1+2\epsilon)}{({\beta y^2_n-1})^2}  }  }.
\label{ratio1}
\end{equation}  
The ratio is a function of $\beta y^2_n$, and 
\begin{equation}
{\gamma_{n(a)}}/{\gamma'_{n} } \approx 1+ 2\epsilon,
\end{equation}
if $\beta y^2_n$ is relatively large.  
\end{enumerate}

\begin{figure}[t]
	\centering
	\includegraphics[width=1.0\linewidth]{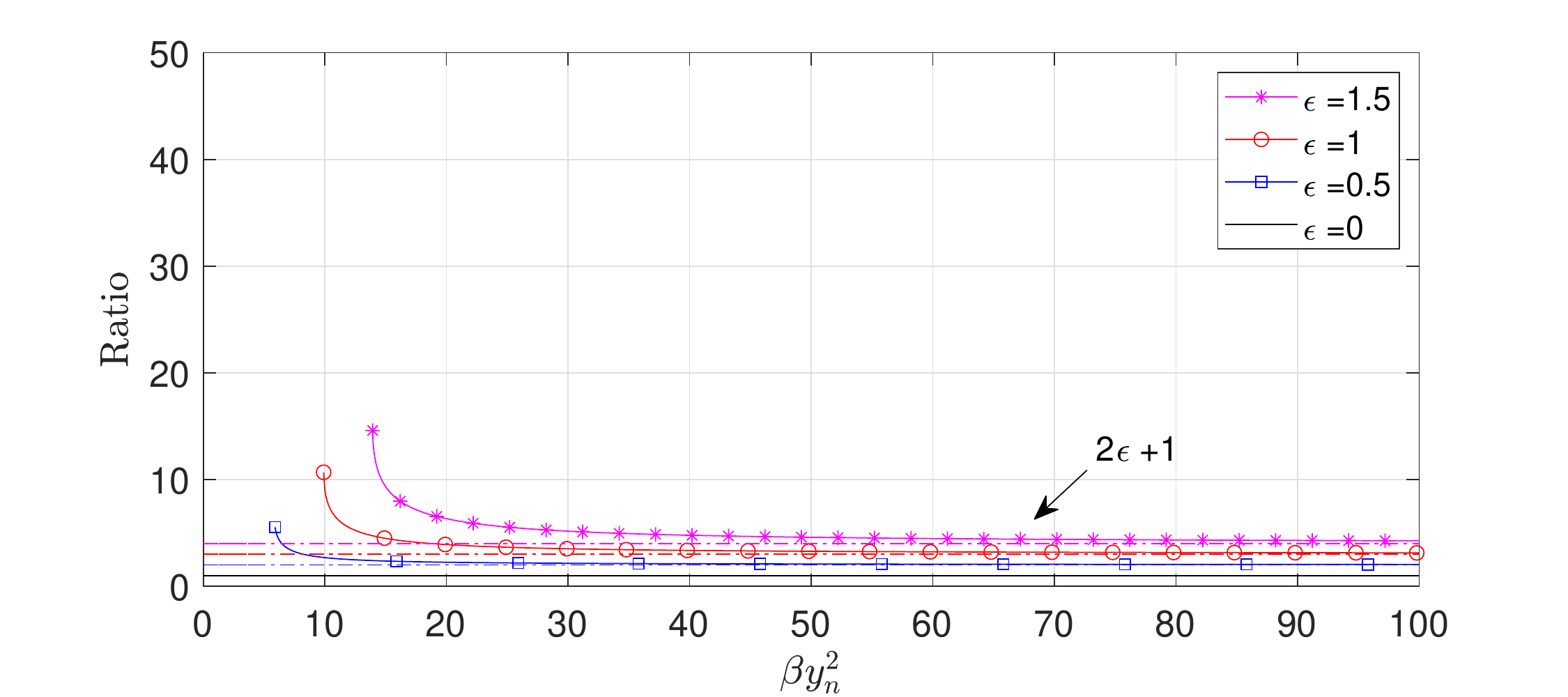}
	\caption{Ratio of precisions with different $\epsilon$.}
	\label{ratio}
\end{figure}
\begin{figure}[t] 
	\centering
	\includegraphics[width=1\linewidth]{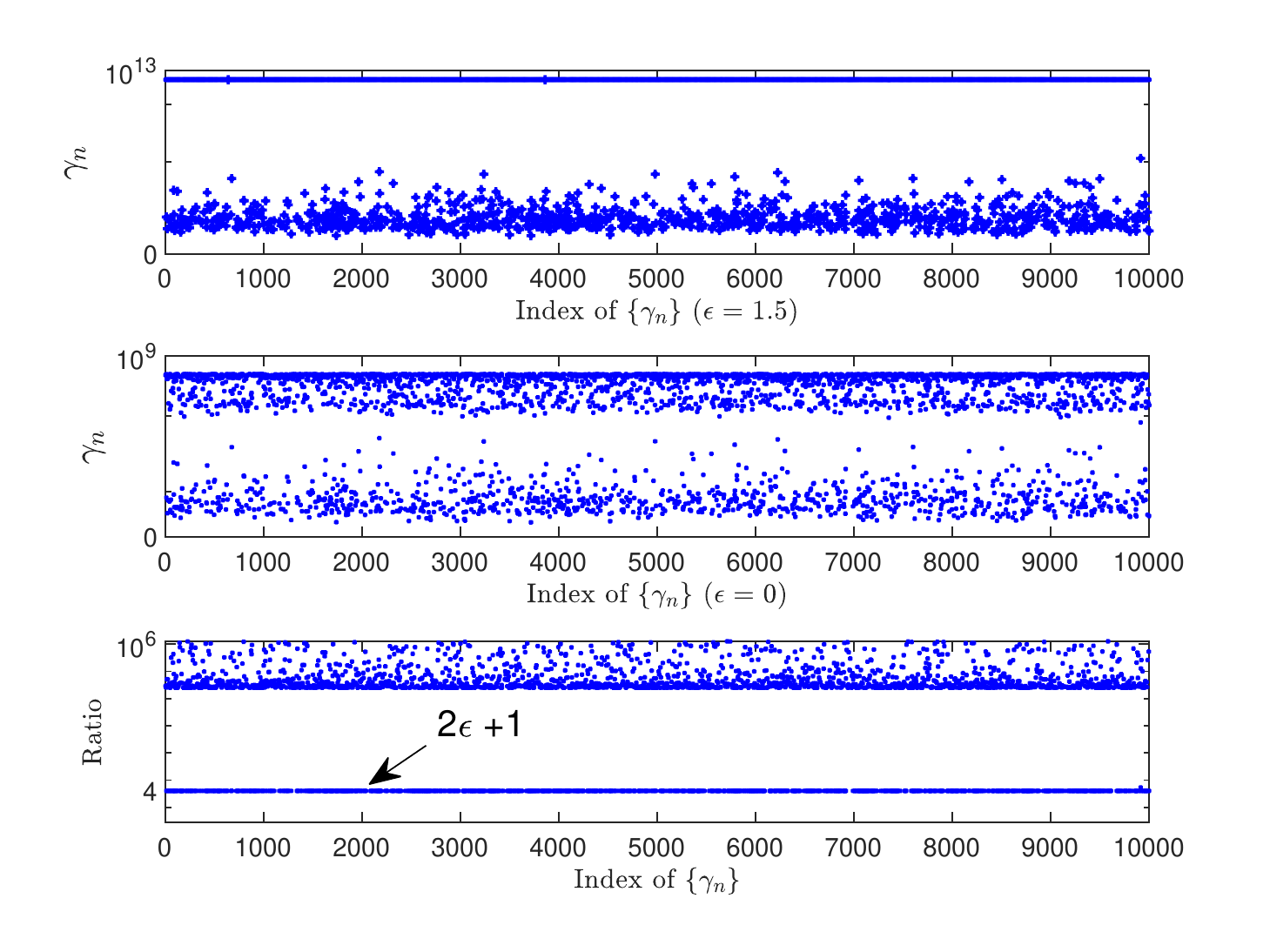}
	\caption{Precisions 
		and their ratios ($\mathbf{A}$ is an identity matrix). }
	\label{demonstrationa}
\end{figure}

\begin{figure}[htbp] 
	\centering
	\includegraphics[width=1\linewidth]{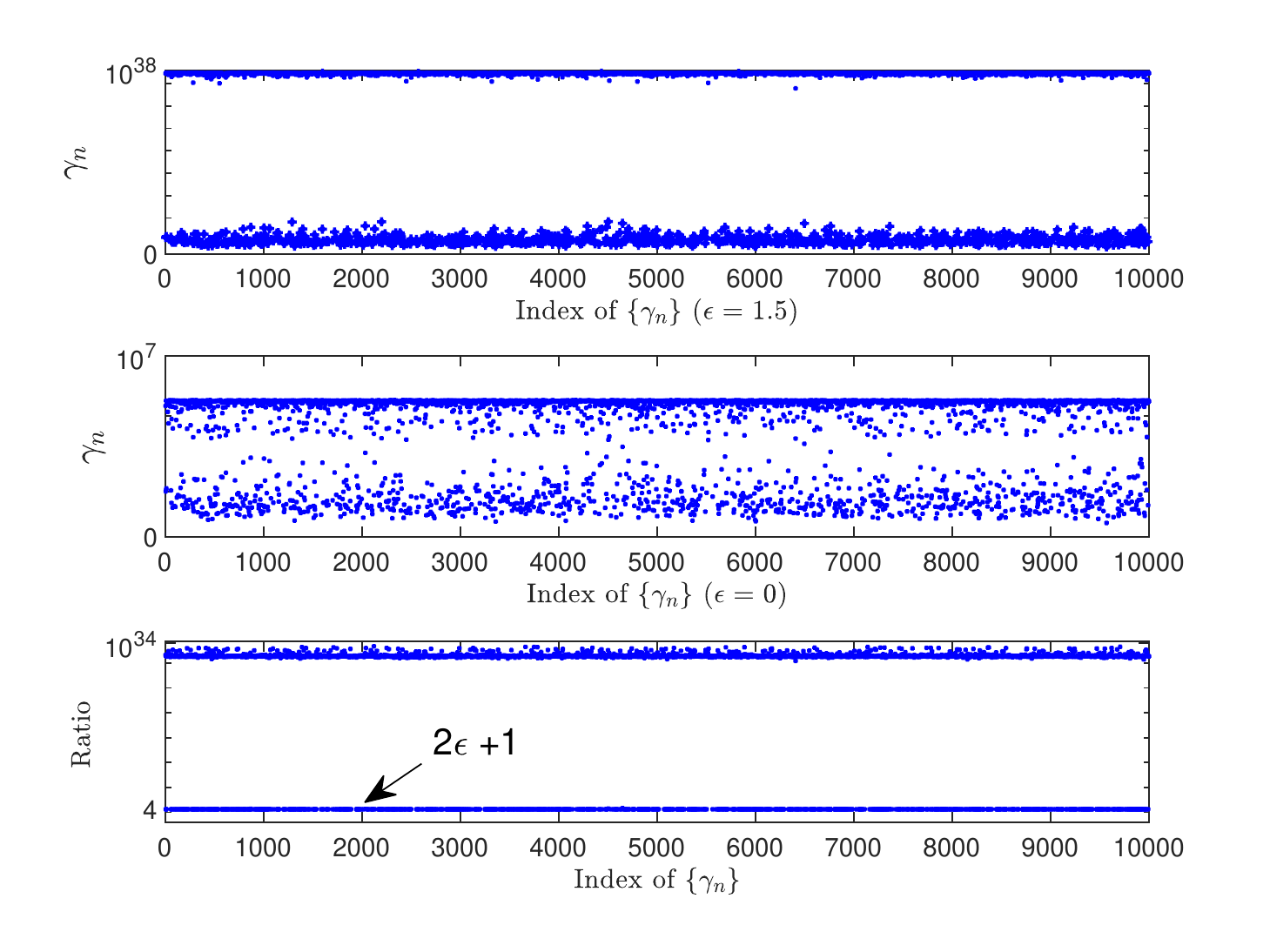}
	\caption{Precisions 
		and their ratios ($\mathbf{A}$ is i.i.d Gaussian).}
	\label{demonstration}
\end{figure}

\rev{The ratios of the precisions versus $\beta y^2_n$ are shown in Fig. \ref{ratio}, where they are not shown for $\beta y^2_n<1+ 4\epsilon +4\sqrt{\epsilon^2 + \epsilon/2}$ as they are infinity when $ 1< \beta y^2_n < 1+ 4\epsilon +4\sqrt{\epsilon^2 + \epsilon/2}$, and undefined when $\beta y^2_n  \leq 1$ (see the above remarks).} It can be seen that the precision obtained with $\epsilon=0$ is amplified depending on the value of $\beta y^2_n$. \rev{The smaller the value of $\beta y^2_n$, the larger the amplification for the corresponding precision (\rea{in} the case of $\beta y^2_n \leq 1$, the ratios are undefined. However, considering a fixed number of iterations, the ratios can be large as $\gamma^t_n$ with a positive $\epsilon$ goes to infinity much quicker).} Note that $y_n=x_n+w_n$ and $\beta$ is the noise precision. Hence, if $\beta y^2_n$ is a small value, it is highly likely that the corresponding $x_n$ is zero, hence the precision $\gamma_n$ should go to infinity. If $\beta y^2_n$ is a large value, it is highly likely that the corresponding $x_n$ is non-zero, hence $\gamma_n$ should be a finite value. \rea{We see that a positive $\epsilon$ tends to a sparser solution, and a proper value of $\epsilon$ leads to much better recovery performance, compared to $\epsilon=0$.} 

The precisions of the elements of the sparse vector obtained by the SBL algorithm with $\epsilon=1.5$ and $\epsilon=0$ are shown in Fig. \ref{demonstrationa}, where \rev{$\mathbf{A}$ is an identity matrix with size $10000 \times 10000$, the sparsity rate of the signal is 0.1, and SNR = 50dB}. It can be seen that the precisions with $\epsilon=1.5$ are separated into two groups more clearly, and the ratios for the small precisions are roughly 4 (i.e., $1+2\epsilon$), while other precisions are amplified significantly.   
Although the above analysis is for an identity matrix $\mathbf{A}$, it is interesting that the same results are observed for a general matrix $\mathbf{A}$ as demonstrated numerically in Fig. \ref{demonstration}, where $\mathbf{A}$ is an i.i.d Gaussian matrix \rev{with size $5000 \times 10000$, the non-zero shape parameter $\epsilon=1.5$, and the sparsity rate and the SNR are the same as the case of identity matrix}. (Similar observations are observed for other matrices). We see that the small precisions are also roughly amplified by 4 times while others are amplified significantly, leading to two well-separated groups.   


\begin{figure}[htbp]
	\centering
	\includegraphics[width=1.1\linewidth]{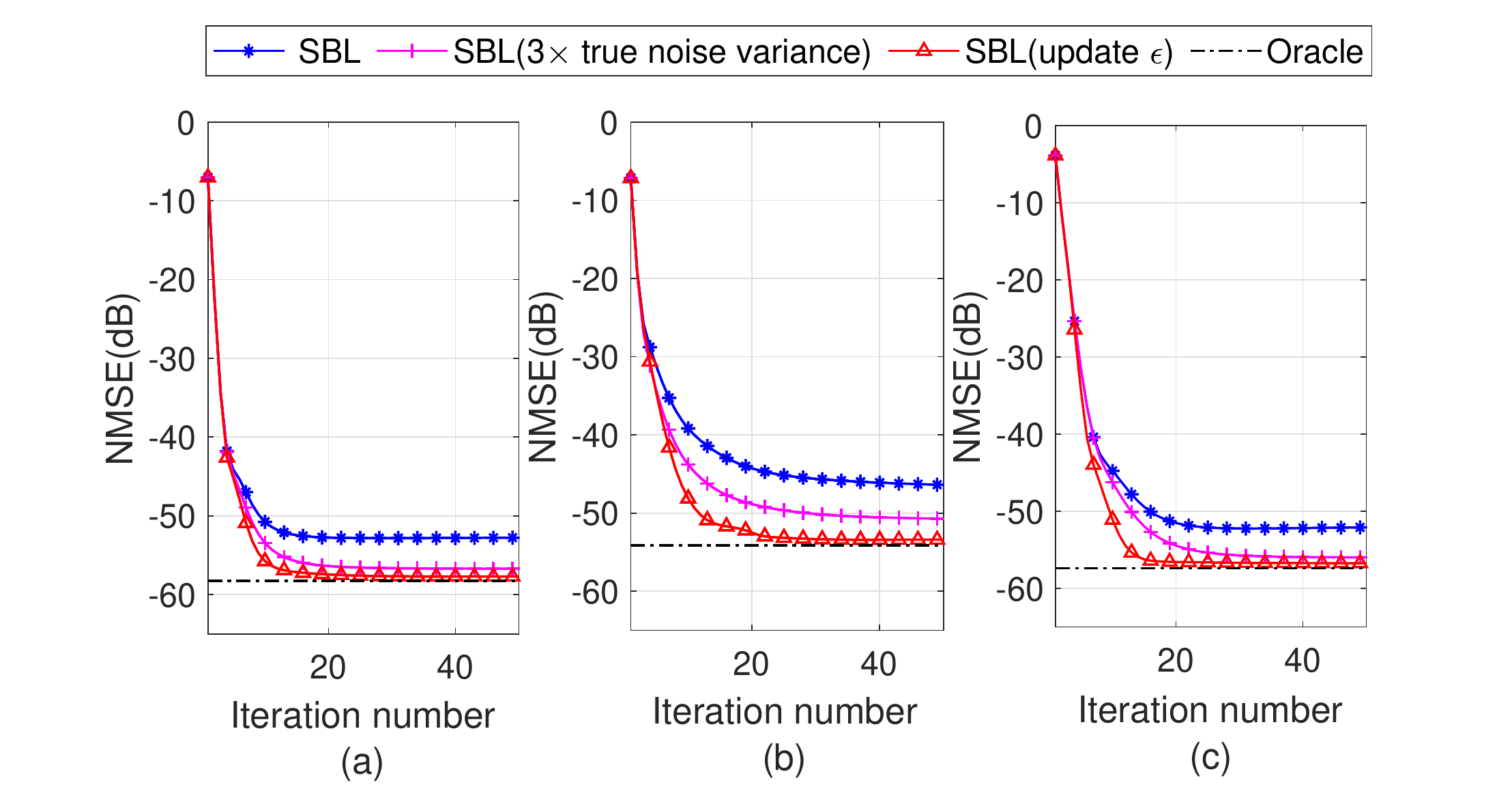}
	\centering
	\caption{Performance of the conventional SBL. (a) Gaussian matrix; (b) correlated matrix with $c=0.3$; (c) low-rank matrix with $ R/N = 0.6$.}
	\label{plugin}
\end{figure}

It is noted that the value of $\epsilon$ should be determined properly. If the matrix $\mathbf{A}$ and the sparsity rate of $\mathbf{x}$ are given, we can find a proper value for $\epsilon$ through trial and error. However, this is inconvenient, and the sparsity rate of the signal may not be available. We found the empirical equation \eqref{eq_eps_update} to determine the value of $\epsilon$. \rea{Next, we examine its effectiveness with the SBL algorithm.}  Plugging the shape parameter update rule (\ref{eq_eps_update}) to the conventional SBL algorithm leads to the following iterative algorithm (assuming the noise precision $\beta$ is known):
\begin{equation}
\begin{aligned}
&Repeat \\
&~~~~~~~~\mathbf{Z}=\left(\beta \mathbf{A}^H\mathbf{A}+Diag(\hat{\bm{\gamma}})\right)^{-1}  \\
&~~~~~~~~\mathbf{\hat x}=\beta \mathbf{ZA}^H\mathbf{y}  \\
&~~~~~~~~\hat\gamma_n=(2 {\epsilon} +1)/(|\hat x_n|^2 + Z_{n,n}), n=1,..., N  \\
&~~~~~~~~ {\epsilon} =
\frac{1}{2}\sqrt{\log(\frac{1}{N}\sum_{n}{\hat{\gamma}_n})-\frac{1}{N}\sum_{n}{\log{\hat{\gamma}}_n}} \nonumber \\
&Until~terminated
\end{aligned}
\end{equation}

To demonstrate the effectiveness of the shape parameter update rule (\ref{eq_eps_update}), we compare the performance of the conventional SBL algorithm with and without shape parameter update. The results are shown in Fig. \ref{plugin}, where the SNR is 50dB, the size of the measurement matrix is $800 \times 1000$, and the {sparsity} rate $\rho = 0.1$. In this figure, the support-oracle bound is also shown for reference. The matrices in (a), (b), and (c) are respectively i.i.d. Gaussian, {correlated and low-rank matrices (refer to \rev{Section VI} for their generations)}. It can be seen that there is a clear gap between the performance of the conventional SBL and the bounds, and with shape parameter updated with our rule, the SBL algorithm attains the bound. It is worth mentioning the empirical finding in \cite{al2018gamp}, i.e., replacing the noise variance $\beta^{-1}$ with $3 \beta^{-1}$ can lead to better performance of GGAMP-SBL \cite{al2018gamp}. We use this for the conventional SBL algorithm and the performance is also included in Fig. \ref{plugin}. \rea{We see that it also leads to substantial performance improvement, but its performance is inferior to that of SBL with updated $\epsilon$ using \eqref{eq_eps_update}}. Moreover, in many cases, the noise variance is unknown, and it may be hard to determine its value accurately. In contrast, our empirical update of $\epsilon$ does not require any additional information.   

\section{Extension to MMV}

In this section, we extend UAMP-SBL to the MMV setting, where the relation among the sparse vectors is exploited, e.g., common support and temporal correlation.   


\subsection{UAMP-SBL for MMV}
The objective on an MMV problem is to recover a collection of length-$N$ sparse vectors $\mathbf{X}=\left[ \mathbf{x}^{(1)},\mathbf{x}^{(2)},...,\mathbf{x}^{(L)} \right]$ from $L$ noisy  length-$M$ measurement vectors $\mathbf{Y}=\left[ \mathbf{y}^{(1)},\mathbf{y}^{(2)},...,\mathbf{y}^{(L)} \right]$ with the following model  
\begin{equation}
\begin{aligned}
\mathbf{Y}= \mathbf{A}  \mathbf{X} + \mathbf{W},
\label{MMV_y=AX+w}
\end{aligned}
\end{equation}
where we assume that the $L$ vectors $\{\mathbf{x}^{(l)}\}$ share a common support (i.e., joint sparsity),  $\mathbf{A}$ is a known measurement matrix with size $M \times N$, and $\mathbf{W}$ denotes an i.i.d. Gaussian noise matrix with the elements having mean zero and precision $\beta$.



\begin{figure}
	\centering
	\includegraphics[width=1\linewidth]{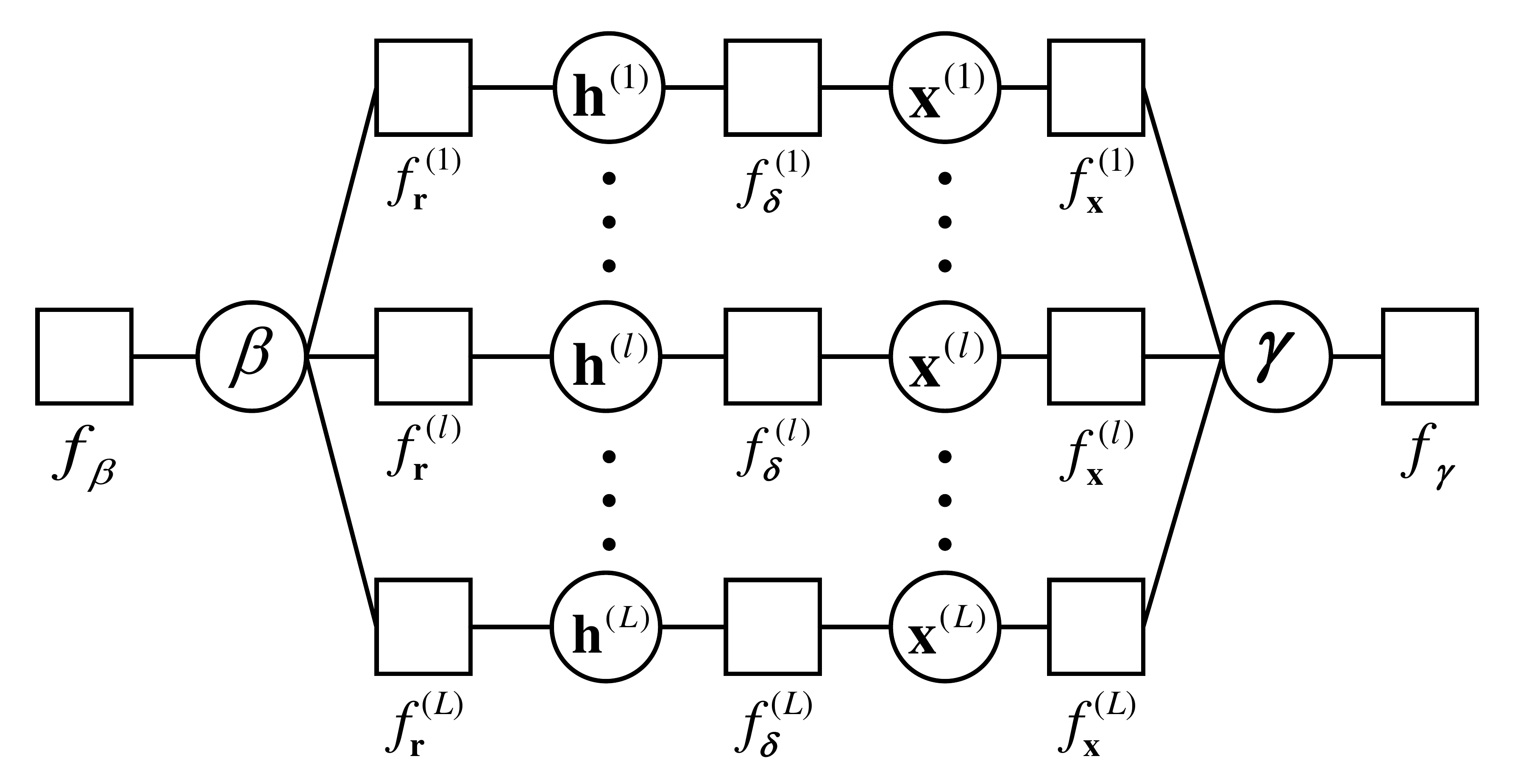}
	\centering
	\caption{Factor graph representation of (\ref{jointPDF_MMV}).}
	\label{fig:factor graph MMV}
\end{figure}

With the SVD  $\mathbf{A=U \Lambda V}$, a unitary transformation with $\mathbf{U}^H$ to (\ref{MMV_y=AX+w}) can be performed, i.e.,
\begin{equation}
\mathbf{R=\Phi X}+\bm{\Omega},
\label{r=uy_MMV}
\end{equation}
where $\mathbf{R=U}^H \mathbf{Y}=\left[ \mathbf{r}^{(1)},\mathbf{r}^{(2)},...,\mathbf{r}^{(L)} \right]$, $\mathbf{\Phi=U}^H \mathbf{A} = \mathbf{\Lambda V}$ and $\bm{\Omega} = \mathbf{U}^H \mathbf{W}$ is still white and Gaussian with mean zero and precision $\beta$. Define {$\mathbf{h}^{(l)}=\mathbf{\Phi x}^{(l)}$} and $\mathbf{H}=[\mathbf{h}^{(1)}, ...,\mathbf{h}^{(L)}]$. Then we have the following joint distribution
\begin{eqnarray}
&& p({\mathbf{X, H}},  \bm{\gamma} ,\beta| {\mathbf{R}}) \nonumber \\  
&&\propto 
\prod_{l=1}^{L}  p( \mathbf{r}^{(l)}|\mathbf{h}^{(l)}, \beta )p( \mathbf{h}^{(l)}| \mathbf{x}^{(l)} ) p(\mathbf{x}^{(l)}|\bm{\gamma} ) p(\bm{\gamma})p(\beta)   \nonumber \\
&&= \prod_{l=1}^{L}  \prod_{m=1}^{M} \mathcal{N} ( {r}_m^{(l)} | {h}_m^{(l)}, {\beta^{-1}})  \delta( {h}_m^{(l)}-[\mathbf{\Phi}]_m  {\mathbf{x}^{(l)}}  ) \nonumber \\
&&\times  \prod_{l=1}^{L} \prod_{n=1}^{N} \mathcal{N} ({x}_n^{(l)} | 0, \gamma_n^{-1})\prod_{n=1}^{N}  \mathrm{Ga}(\gamma_n|\epsilon,\eta) p(\beta) .
\label{jointPDF_MMV}
\end{eqnarray}
Define factors 
$f_{\mathbf{r}}^{(l)}({\mathbf{r}^{(l)}}, \mathbf{h}^{(l)}, \beta ) =\prod_m \mathcal{N}( {r}_m^{(l)}|{h}_m^{(l)}, \beta)$, $ f_{\bm{\delta}}^{(l)} (\mathbf{h}^{(l)},\mathbf{x}^{(l)})  =  \prod_m \delta({h}_m^{(l)}-[\mathbf{\Phi}]_m \mathbf{x}^{(l)} ) $,
$f_{\beta}(\beta) \propto 1/\beta$,   
$ f_{\mathbf{x}}^{(l)}(\mathbf{x}^{(l)},\bm{\gamma}) =
\prod_n \mathcal{N} (x_n^{(l)} | 0, \gamma_n^{-1} )$, and $f_{\bm{\gamma}} (\bm{\gamma},\epsilon) =\prod_n \mathrm{Ga}({\gamma_n}|\epsilon,\eta) $ denotes the hyperprior of the hyperparameters $\{\gamma_n\} $. The factor graph representation of (\ref{jointPDF_MMV}) is shown in Fig.~\ref{fig:factor graph MMV}\footnote{\rev{The vector variable node $\bm \gamma$ is used in the factor graph to make it neat. We note that each entry $x^{(l)}_n$ of $\bm{x}^{(l)}$ is connected to $\gamma_n$ through the function node between them.}}, based on which the message passing algorithm can be derived.  The message updates related to $\mathbf{x}^{(l)}$ and $\mathbf{h}^{(l)}$ are the same as those for the SMV case and can be computed in parallel. The difference lies in the computations of $\hat\beta$ and  $\hat{\bm\gamma}$, \rev{and the relevant derivations are shown in Appendix D.} The UAMP-SBL for MMV is summarized in Algorithm \ref{UTAMP_SBL_MMV-table}, \rev{where UAMPv2 is employed. The complexity of the algorithm is $\mathcal{O} (MNL)$ per iteration.}

\begin{algorithm}\small
	\caption{UAMP-SBL for MMV}
	Unitary transform: $\mathbf{R=U}^H \mathbf{Y}=\mathbf{\Phi} {\mathbf{X}} +{\mathbf{W}}$, where $\mathbf{\Phi=U}^H\mathbf{A}=\mathbf{\Lambda V}$, and $\mathbf{A}$ has SVD $\mathbf{A=U \Lambda V}$.\\
	Define vector $\bm{\lambda}= \mathbf{\Lambda \Lambda}^H \mathbf{1}$.\\  
	Initialization: $\forall l$; ${\tau}_x^{l(0)}=1$, $\hat{\mathbf{x}}^{l(0)}=\textbf{0}$, {${\epsilon'}=0.001$}, $\hat{{\bm{\gamma}}}=\textbf{1}$, $\hat{ \beta}=1$, $\mathbf{s}^{l}=\mathbf{ 0 }$, and $t=0$.\\
	\textbf{Do}     
	\begin{algorithmic}[1]
		\STATE $\forall l$;  $ 	 	{\bm{\tau}_p}^l$ = $ \tau^{l(t)}_x  \bm{\lambda}$\\
		\STATE $\forall l$; $ 	 	\mathbf{p}^l=\mathbf{\Phi} \hat{\mathbf{x}}^{l(t)} - \bm{\tau}_{p}^{l} \cdot  \mathbf{s}^{l} $\\
		\STATE $\forall l$;  ${\mathbf{v}^l_h}  = \bm{\tau}_p^{l}./ (\bm{1}+\hat{ \beta}\bm{\tau}_p^{l} )$\\  
		\STATE $\forall l$; $  \mathbf{\hat{h}}^l = (\hat\beta\bm{\tau}_{p}^{l}\cdot \mathbf{r}^{l}+\mathbf{p}^{l})./(\bm{1}+\hat\beta\bm{\tau}_p^{l}) $\\
		\STATE $\hat{ \beta} =   {LM}/   ( \sum_l ( {  ||  \mathbf{r}^{l}-  \mathbf{\hat{h}}^{l}  ||^2   + \mathbf{1}^H\mathbf{v}_h ^{l} } ) ) $;\\	
		\STATE $\forall l$;  $     \bm{\tau}_s^{l} = \mathbf{1}./ (\bm{\tau}_p^{l}+  {\hat{\beta}}^{-1}\mathbf{1}) $\\
		\STATE$\forall l$;  $      \mathbf{s}^l= \bm{\tau}^{l}_s \cdot (\mathbf{r}^{l}-\mathbf{p}^{l}) $\\
		\STATE 	$\forall l$; $    1/\tau_q^{l} = ({1}/{N}) \bm{   \lambda }^H \bm{\tau_s   }^{l}$\\
		\STATE$\forall l$;  $     \mathbf{q}^{l} =\hat{\mathbf{x}}^{l(t)} + \tau_q^{l} (\mathbf{\Phi}^H  \mathbf{s}^{l})$\\
		\STATE 	$\forall l$; $   \tau_x^{l(t+1)}=  (\tau_q^{l}/N) \mathbf{1}^H(\mathbf{1}. /(\mathbf{1}+\tau_q^{l}  \hat{{\bm{\gamma}}} ))$
		\STATE 	$\forall l$; $   \hat{\mathbf{x}}^{l(t+1)}= 	 \mathbf{q}^l   ./(\mathbf{1}+\tau_q^l \hat{{\bm{\gamma}}} )$\\
		\STATE  $\hat{{\gamma}}_n = \frac{2{\epsilon'} + 1}{ (1/L) \sum^L_{l=1} (|{{\hat{x}}_n^{l(t+1)}}|^2 +\tau_x^{l(t+1)})} , n=1, ...,N.$ \\
		\STATE ${\epsilon'}=\frac{1}{2}\sqrt{\log(\frac{1}{N}\sum_{n}{ {\hat{\gamma}_n}})-\frac{1}{N}\sum_{n}{\log{{\hat{\gamma}_n}}}}$ \\
		\STATE   $ t=t+1 $  
	\end{algorithmic}
	\textbf{while}	 {$ \frac{1}{L} \sum_{l=1}^L  ( || \hat{\mathbf{x}}^{l(t+1)} - \hat{\mathbf{x}}^{l(t)} ||^2/ ||  \hat{\mathbf{x}}^{l(t+1)}  ||^2  ) > \delta_x$} and $t<t_{max}$)
	\label{UTAMP_SBL_MMV-table}
\end{algorithm}

\subsection{UAMP-TSBL}
With the assumption of a common sparsity profile shared by all sparse vectors, we further consider exploiting the temporal correlation that exists between the non-zero elements. 
The messages update related to  $\mathbf{h}^{(l)}$, ${\epsilon}$ and $\beta$ are the same as those for the MMV case, where no temporal correlation between non-zero elements is assumed. As the correlation is considered, the differences from the UAMP-SBL MMV algorithm lie in the computations of $\hat{\gamma}_n$ and  $\mathbf{x}^{(l)}$. 

\begin{figure}
	\centering
	\includegraphics[width=0.75\linewidth]{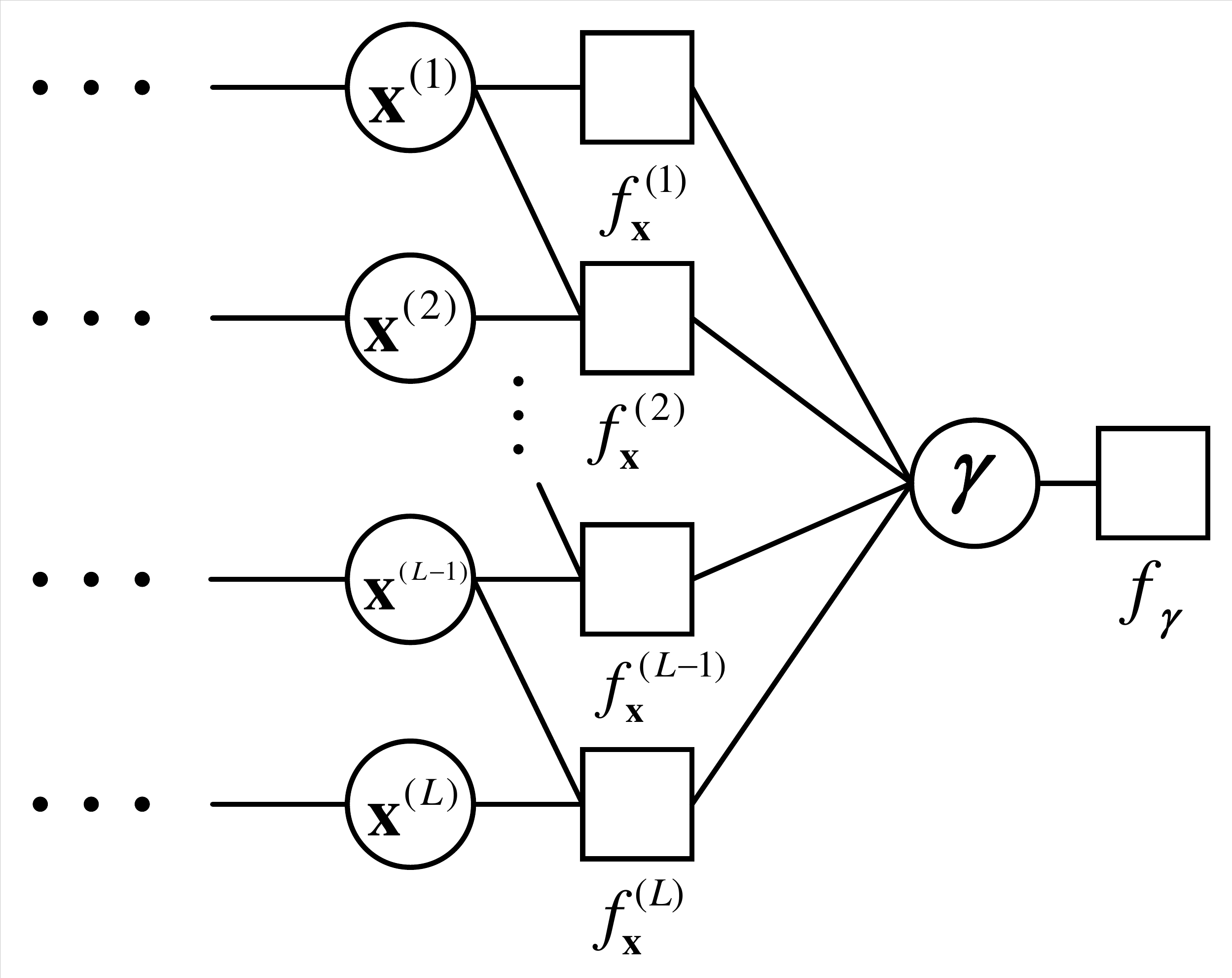}
	\centering
	\caption{\rev{The additional factor graph for deriving UAMP-TSBL.}}
	\label{fig:figtsbl}
\end{figure}

As in {\cite{al2018gamp}}, we use an AR(1) process  \cite{zhang2010sparse} to model the correlation between $x^{(l)}_n$ and $x^{(l-1)}_n$, i.e., 
\begin{equation}
\begin{aligned}
x^{(l)}_n&=\alpha x^{(l-1)}_n + \sqrt{1-\alpha^2} \vartheta^{(l)}_n \\
p(x^{(l)}_n|x^{(l-1)}_n)&=\mathcal{N} (x^{(l)}_n|\alpha x^{(l-1)}_n,(1-\alpha^2) {{\gamma}^{-1}_n}), l >1 \\
p(x^{(1)}_n)&=\mathcal{N} (x^{(1)}_n|0, {{\gamma}^{-1}_n}),
\end{aligned}
\end{equation}
where  $\alpha \in (-1,1)$ is the temporal correlation coefficient and {$\vartheta^{(l)}_n \sim \mathcal{N}(0,{{\gamma}^{-1}_n})$}. Due to the temporal correlation, the conditional prior distribution for the vector $\mathbf{x}^{(l)}$ changes. We redefine the factors $\{ f_{x^{(l)}_n} (x^{(l)}_n, {\gamma_n}) \}$, i.e.,   $f_{x^{(l)}_n}(x^{(l)}_n,{\gamma_n})= p(x^{(l)}_n|x^{(l-1)}_n)$ for $l>1$ and $f_{x^{(1)}_n}(x^{(1)}_n, {\gamma_n})=p(x^{(1)}_n)$. Thus, 
each $x_n^{(l)}$ is connected to the factor nodes $f^{(l )}_{x_n}(x^{(l)}_n|{\gamma_n})$, $f^{(l+1)}_{x_n}(x^{(l+1)}_n|{\gamma_n})$  and $\{f^{(l)}_{\delta_m}(h^{(l)}_m| \mathbf{x}^{(l)}) ,\forall m \}$. \rev{\rea{The factor graph characterizing the temporal correlation} is shown in Fig. \ref{fig:figtsbl}. The remaining part of the graph is omitted as it is the same as that of the MMV case without temporal correlation. The derivation of the extra message passing for the UAMP-TSBL algorithm is shown in Appendix E, and the algorithm is summarized in Algorithm \ref{UTAMP_TSBL-table}. UAMP-TSBL is an extension of the UAMP-SBL algorithm for MMV (Algorithm \ref{UTAMP_SBL_MMV-table}). 
The complexity of the UAMP-TSBL algorithm is also dominated by matrix-vector multiplications, and it is $\mathcal{O} (MNL)$ per iteration.   } 

\begin{algorithm}\small
	\caption{UAMP-TSBL}
	Unitary transform: $\mathbf{R=U}^H \mathbf{Y}=\mathbf{\Phi} {\mathbf{X}} +\mathbf{W}$, where $\mathbf{\Phi=U}^H\mathbf{A}=\mathbf{\Lambda V}$, and $\mathbf{A}$ has SVD $\mathbf{A=U \Lambda V}$.\\
	Define vector $\bm{\lambda}= \mathbf{\Lambda \Lambda}^H \mathbf{1}$.\\  
	Initialization: $\forall l$: $ {\tau}_x^{l(0)}={1}$, $\hat{\mathbf{x}}^{l(0)}=\mathbf{ 0 }$, $\mathbf{q}^{l}={\mathbf{0}}$,${\bm{\tau}_q^{{l(0)}}}=\mathbf{1}$, $\bm{\xi}^{l(0)}=\mathbf{ 0 }$,  $\bm{\psi}^{l(0)}={\mathbf{1}}$, 
	$\bm{\theta}^{l(0)}=\mathbf{ 0 }$,  $\bm{\phi}^{l(0)}={\mathbf{1}}$, $\mathbf{s}^{l(-1)}=\mathbf{0}$,  ${\epsilon'}=0.001$, $\hat{{\bm{\gamma}}}^{(0)}=\textbf{1}$, $\hat{ \beta}=1$, and $t=0$.\\
	\textbf{Do}     
	\begin{algorithmic}[1]
		\STATE $  \bm{\xi}^{1} = \mathbf{0}$\\
		\STATE $  \bm{\psi}^{1} = \mathbf{1}./\hat{\bm{\gamma}}^{(t)}$\\	
		\STATE $\textbf{for $l = 2,...,L$}$\\
		\STATE $\quad \bm{\xi}^{l} = \alpha \left( \frac{\mathbf{q}^{l-1} }{\bm{\tau}_q^{l-1}}+  \frac{\bm{\xi}^{l-1} }{\bm{\psi}^{l-1}}    \right) \cdot \left( \frac{\bm{\tau}_q^{l-1}\cdot \bm{\psi}^{l-1} }{\bm{\tau}_q^{l-1} + \bm{\psi}^{l-1}}  \right)   $\\
		\STATE $\quad \bm{\psi}^{l} = \alpha^2   \left( \frac{\bm{\tau}_q^{l-1} \cdot \bm{\psi}^{l-1} }{\bm{\tau}_q^{l-1} + \bm{\psi}^{l-1}}  \right) + (1-\alpha^2)/  \hat{\bm{\gamma}}^{(t)}  $\\
		\STATE $\textbf{end}$\\
		
		\STATE $\textbf{for $l = 1,...,L$}$\\
		\STATE   $ \quad	 	{\bm{\tau}_p}^l$ = $ \tau^{l(t)}_x  \bm{\lambda}$\\
		\STATE   $ \quad	 	\mathbf{p}^l=\mathbf{\Phi} \hat{\mathbf{x}}^{l(t)} - \bm{\tau}_{p}^{l} \cdot  \mathbf{s}^{l(t-1)} $\\
		\STATE    $ \quad       {\mathbf{v}^l_h}  = \bm{\tau}_p^{l}./ (\bm{1}+\hat{ \beta}\bm{\tau}_p^{l} )$\\  
		
		\STATE   $  \quad       \mathbf{\hat{h}}^l = (\hat\beta\bm{\tau}_{p}^{l}\cdot \mathbf{r}^{l}+\mathbf{p}^{l})./(\bm{1}+\hat\beta\bm{\tau}_p^{l}) $\\
		\STATE $\textbf{end}$\\
		\STATE $\hat{ \beta} =   {LM}/   ( \sum_l ({  ||  \mathbf{r}^{l}-  \mathbf{\hat{h}}^{l}  ||^2   + \mathbf{1}^H\mathbf{v}_h ^{l} } ) ) $\\	
		\STATE $\textbf{for $l = 1,...,L$}$\\
		
		\STATE   $  \quad   \bm{\tau}_s^{l} = \mathbf{1}./ (\bm{\tau}_p^{l}+  \hat{ \beta}^{-1}\mathbf{1}) $\\
		\STATE   $  \quad    \mathbf{s}^{l(t )}= \bm{\tau}^{l}_s \cdot (\mathbf{r}^{l}-\mathbf{p}^{l}) $\\
		\STATE   $  \quad  1/\tau_q^{l} = ({1}/{N}) \bm{   \lambda }^H \bm{\tau_s   }^{l}$\\
		\STATE $  \quad   \mathbf{q}^{l} =\hat{\mathbf{x}}^{l(t)} + \tau_q^{l} (\mathbf{\Phi}^H  \mathbf{s}^{l(t)})$\\	
		
		\STATE 	$ \quad   {\tau}_x^{l (t+1)}=  (1/N) \mathbf{1}^H(\mathbf{1}. /(\mathbf{1}./ \bm{\tau}^{l}_q +\mathbf{1}./\bm{\phi}^{l}+\mathbf{1}./\bm{\psi}^{l}))$
		\STATE 	$ \quad  \hat{\mathbf{x}}^{l (t+1)} =  {\tau}_x^{l (t+1)} (\mathbf{q}^{l}./ \bm{\tau}^{l}_q +\bm{\theta}^{l}./\bm{\phi}^{l}+\bm{\xi}^{l}./\bm{\psi}^{l})  $\\
		\STATE $\textbf{end}$\\

		\STATE $  \bm{\theta}^{L-1} =  \frac{1}{\alpha}\mathbf{q}^{L} $\\
		\STATE $  \bm{\phi}^{L-1} = \frac{1}{\alpha^2}   \left(  \bm{\tau}_q^{L} + (1-\alpha^2)/\hat{\bm{\gamma}}^{(t)} \right) $\\
		
		\STATE $\textbf{for $l = L-2,...,1$}$\\
		\STATE $\quad \bm{\theta}^{l} = \frac{1}{\alpha} \left( \frac{\mathbf{q}^{l+1} }{\bm{\tau}_q^{l+1}}+  \frac{\bm{\theta}^{l+1} }{\bm{\phi}^{l+1}}    \right)  \cdot \left( \frac{\bm{\tau}_q^{l+1} \bm{\phi}^{l+1} }{\bm{\tau}_q^{l+1} + \bm{\phi}^{l+1}}  \right)   $\\
		\STATE $\quad \bm{\phi}^{l} = \frac{1}{\alpha^2}   \left( \frac{\bm{\tau}_q^{l+1} \bm{\phi}^{l+1} }{\bm{\tau}_q^{l+1} + \bm{\phi}^{l+1}}  + (1-\alpha^2)/ \hat{\bm{\gamma}}^{(t)} \right)  $\\
		\STATE $\textbf{end}$\\
		
		\STATE $ \hat{\bm{\gamma}}^{(t+1)}=  L( {2 {\epsilon'} +1})/[|{\mathbf{\hat{x}}^{1(t+1)}}|^2
		+ {\tau}_x^{1(t+1)}\mathbf{1}  $\\ $+\frac{1}{1-\alpha^2}\sum_{l=2}^{L}( |{\mathbf{\hat{x}}^{l(t+1)}}|^2 + {\tau}_x^{l(t+1)}\mathbf{1} )$\\
		$+\frac{\alpha^2}{1-\alpha^2}\sum_{l=1}^{L-1}( |{\mathbf{\hat{x}}^{l(t+1 )}}|^2 + {\tau}_x^{l(t +1)}\mathbf{1} ) - \frac{2\alpha}{1-\alpha^2} \sum_{l=2}^{L}( {\mathbf{\hat{x}}^{l(t+1)}}\cdot {\mathbf{\hat{x}}^{(l-1)(t+1 )}} )] $ \\
		\STATE${\epsilon'}=\frac{1}{2}\sqrt{\log(\frac{1}{N}\sum_{n}{ {\hat{\gamma}^{(t+1)}_n}})-\frac{1}{N}\sum_{n}{\log{{\hat{\gamma}_n}} ^{(t+1)}}}$\\  
		\STATE $t=t+1$  
	\end{algorithmic}
	\textbf{while}	 {$ \frac{1}{L} \sum_{l=1}^L  (  | |  \hat{\mathbf{x}}^{l(t+1)}-   \hat{\mathbf{x}}^{l(t)}| |^2/ | |    \hat{\mathbf{x}}^{l(t+1)} | |^2  ) > \delta_x$} and $t<t_{max}$ 
	\label{UTAMP_TSBL-table}
\end{algorithm}


\section{Numerical results}
\label{sec:Numerical results}

In this section, we compare the proposed UAMP-(T)SBL algorithms with the conventional SBL and state-of-the-art AMP-based SBL algorithms. \rea{We evaluate the performance of various algorithms using normalized MSE,  defined as}
\begin{eqnarray}
\text{NMSE} &\triangleq& \frac{1}{\rev{K}}\sum_{k=1}^K {  || \hat{ \mathbf{x}}_{k}- { \mathbf{x}_{k}}  || ^2}/ { || { \mathbf{x}_{k}} || ^2}, \label{NMSE} \\
\text{NMSE} &\triangleq& \frac{1}{\rev{K}{L}}\sum_{k=1}^J \sum_{l=1}^L {|| \hat{\mathbf{x}}^{(l)}_k -  {\mathbf{x}}^{(l)}_k||^2}/ {||   {\mathbf{x}}^{(l)}_k ||^2} 
\end{eqnarray}
for the SMV and MMV cases respectively, where \rev{ $\hat{ \mathbf{x}}_{k}~ (\hat{ \mathbf{x}}_{k}^{(l)})$ is the estimate of { ${\mathbf{x}}_{k}~({ \mathbf{x}}_{k}^{(l)})$}, and $\rev{K}$} is the number of trials. Since different algorithms have different computational complexity per iteration and they require a different number of iterations to converge, as in \cite{al2018gamp}, we measure the runtime of the algorithms to indicate their relative computational complexity. \rev{It is noted that the time consumed by the SVD in UAMP-SBL is counted \rea{for the runtime.}} 
\begin{figure}[t]
	\centering
	\includegraphics[width=1.0\linewidth]{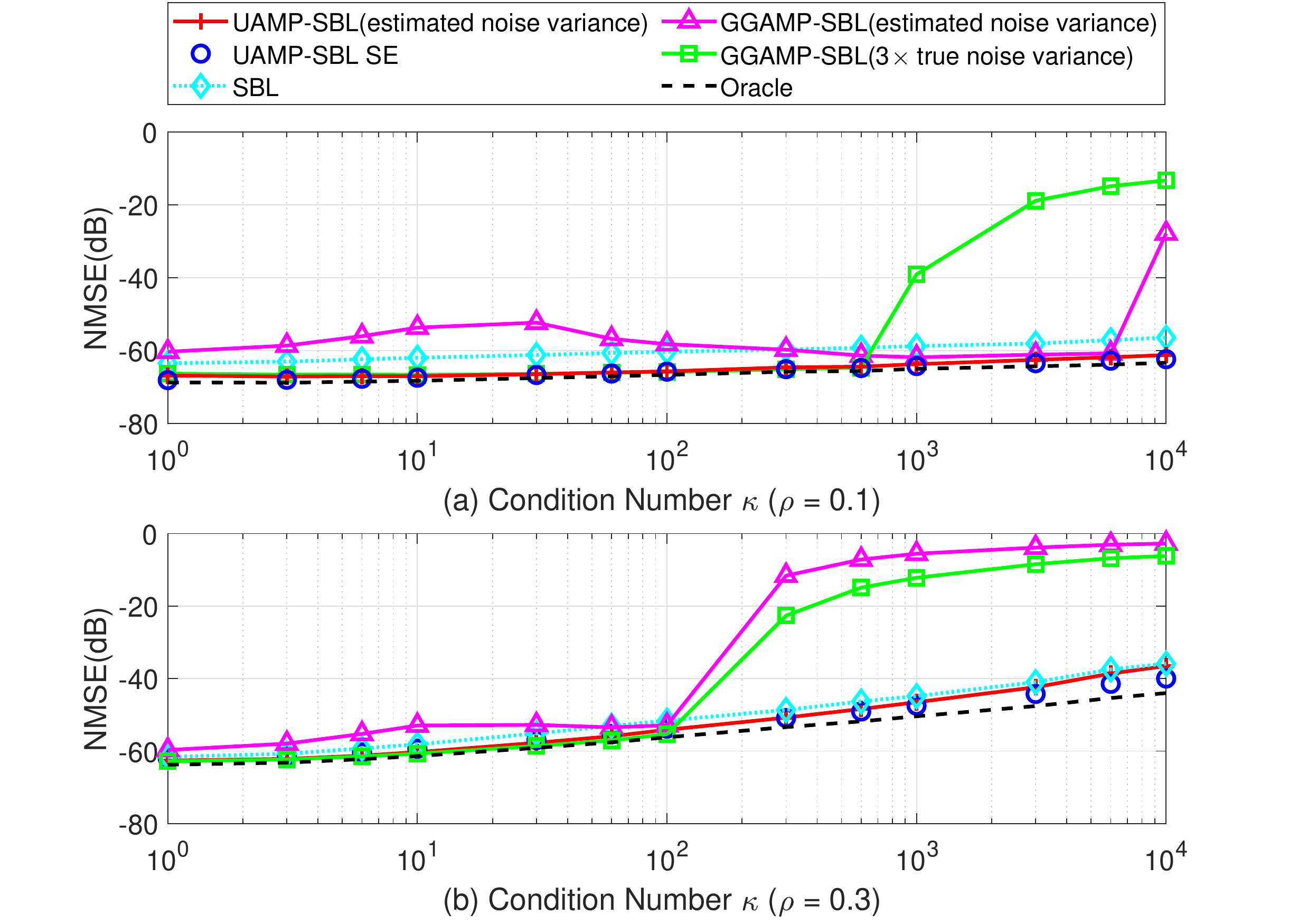}	\centering
	\caption{Performance comparison (ill-conditioned matrices).}
	\label{conditionresult}
\end{figure}
\begin{figure}[htbp]
	\centering
	\includegraphics[width=1.0\linewidth]{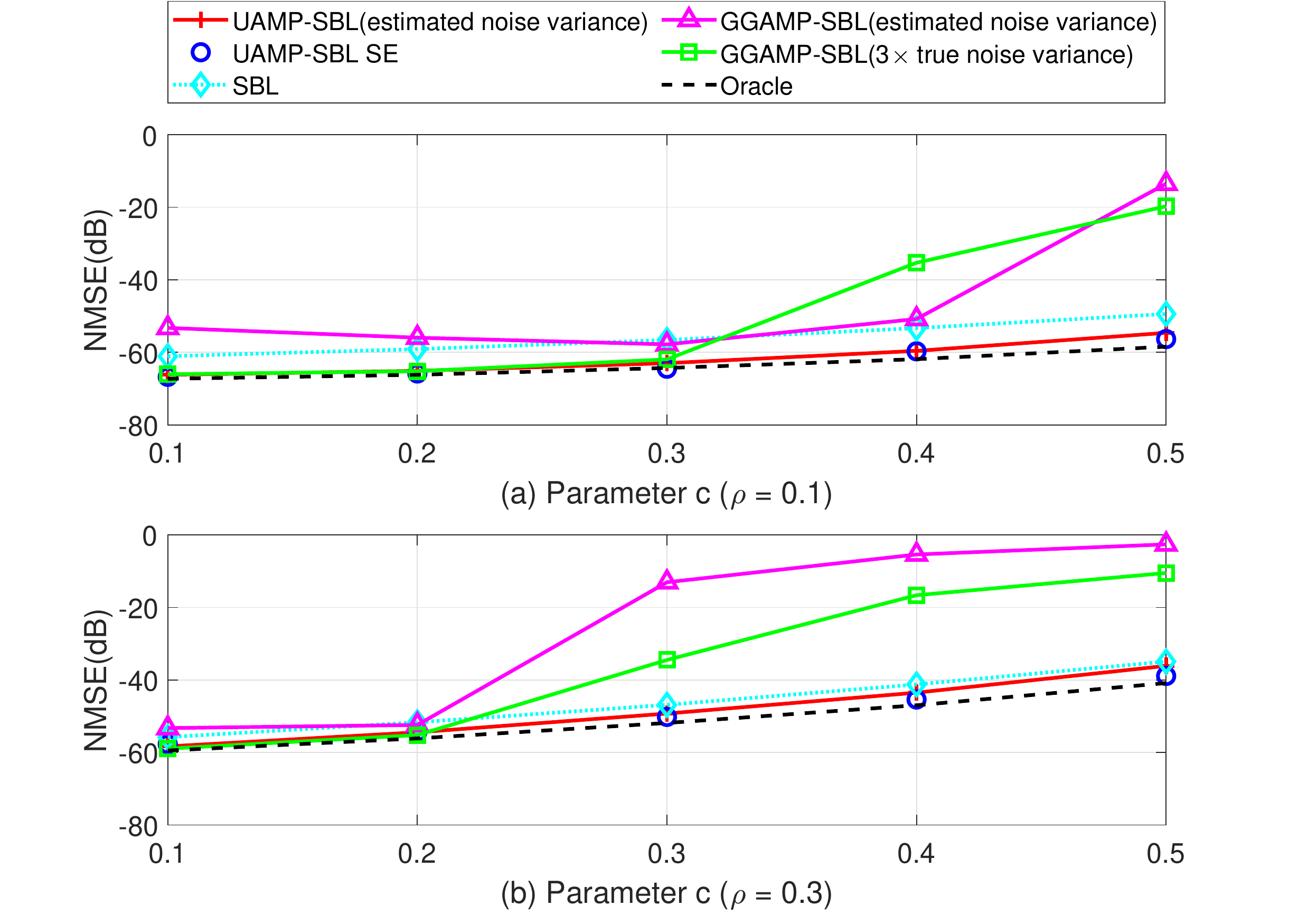}
	\caption{Performance comparison (correlated matrices).}
	\label{correlationresult}
\end{figure}

To test the robustness and performance of the algorithms, we use the following measurement matrices: 
\begin{enumerate}
	\item Ill-conditioned Matrix: Matrix $\mathbf{ A }$ is constructed based on the SVD $\mathbf{A}=\mathbf{U \Lambda V}$ where $\bm{\Lambda}$ is a singular value matrix with $\Lambda_{i,i}/ \Lambda_{i+1,i+1} = \kappa^{1/(M-1)}$ { for $i=1,2,...,M-1$ } (i.e., the condition number of the matrix is $\kappa$).
	\item Correlated Matrix: The correlated matrix $\textbf{A}$ is constructed using  $\textbf{A}=\textbf{C}_L^{1/2}\textbf{G}\textbf{C}_R^{1/2}$, where $\textbf{G}$ is an { i.i.d. }  Gaussian matrix with mean zero and unit variance, and $\textbf{C}_L$ is an $M\times M$ matrix with the $(m,n)$th element given by $c^{|m-n|}$ where $c\in[0,1]$. Matrix $\textbf{C}_R$ is generated in the same way but with a size of $N\times N$. The parameter $c$ controls the correlation of matrix $\textbf{A}$. 
	\item Non-zero Mean Matrix: The elements of matrix $\mathbf{A}$ are drawn from a non-zero mean Gaussian distribution, i.e., $a_{m,n} \sim  \mathcal{N}( a_{m,n} | \mu,  1)$. The mean $\mu$ measures the derivation from the {i.i.d.} zero-mean Gaussian matrix.
	\item Low Rank Matrix: The measurement matrix $\textbf{A} =\textbf{BC}$, where the size of $\textbf{B}$ and $\textbf{C}$ are $M\times R$ and $R\times N$, respectively, and $R<M$. Both $\textbf{B}$ and $\textbf{C}$
	are i.i.d. Gaussian matrices with mean zero and unit variance. The rank ratio $R/N$ is used to measure the deviation of matrix $\textbf{A}$ from the i.i.d. Gaussian matrix.  
\end{enumerate}

\subsection{Numerical Results for SMV}
In this section, we compare UAMP-SBL  against the   {conventional }SBL {\cite{tipping2001sparse}} and  the state-of-the-art AMP based SBL algorithm GGAMP-SBL \cite{al2018gamp} with estimated noise variance and 3 times of the true noise variance. 
The vector $\mathbf{x}$ is drawn from a Bernoulli-Gaussian distribution with a non-zero probability $\rho$. The SNR is defined as $\text{SNR} \triangleq E\left\| \mathbf{ \mathbf{A x}} \right\| ^2/ E\left\| \mathbf{ \mathbf{w}} \right\| ^2$. As a performance benchmark, the support-oracle MMSE bound \cite{al2018gamp} is also included. We set $M=800$, $N=1000$ and \rev{the SNR is set to be 60dB, unless it is specified}. For UAMP-SBL we set the maximum iteration number $t_{max}=300$ (note that there is no inner iteration in UAMP-SBL). GGAMP-SBL is a double loop algorithm, the maximum numbers of E-step and outer iteration are set to be 50 and 1000 respectively. The damping factor for GGAMP-SBL is 0.2 to enhance its robustness against tough measurement matrices. It is noted that the damping factor can be increased to reduce the runtime of GGAMP-SBL but at the cost of reduced robustness.

\begin{figure}[t]
	\centering
	\includegraphics[width=1.0\linewidth]{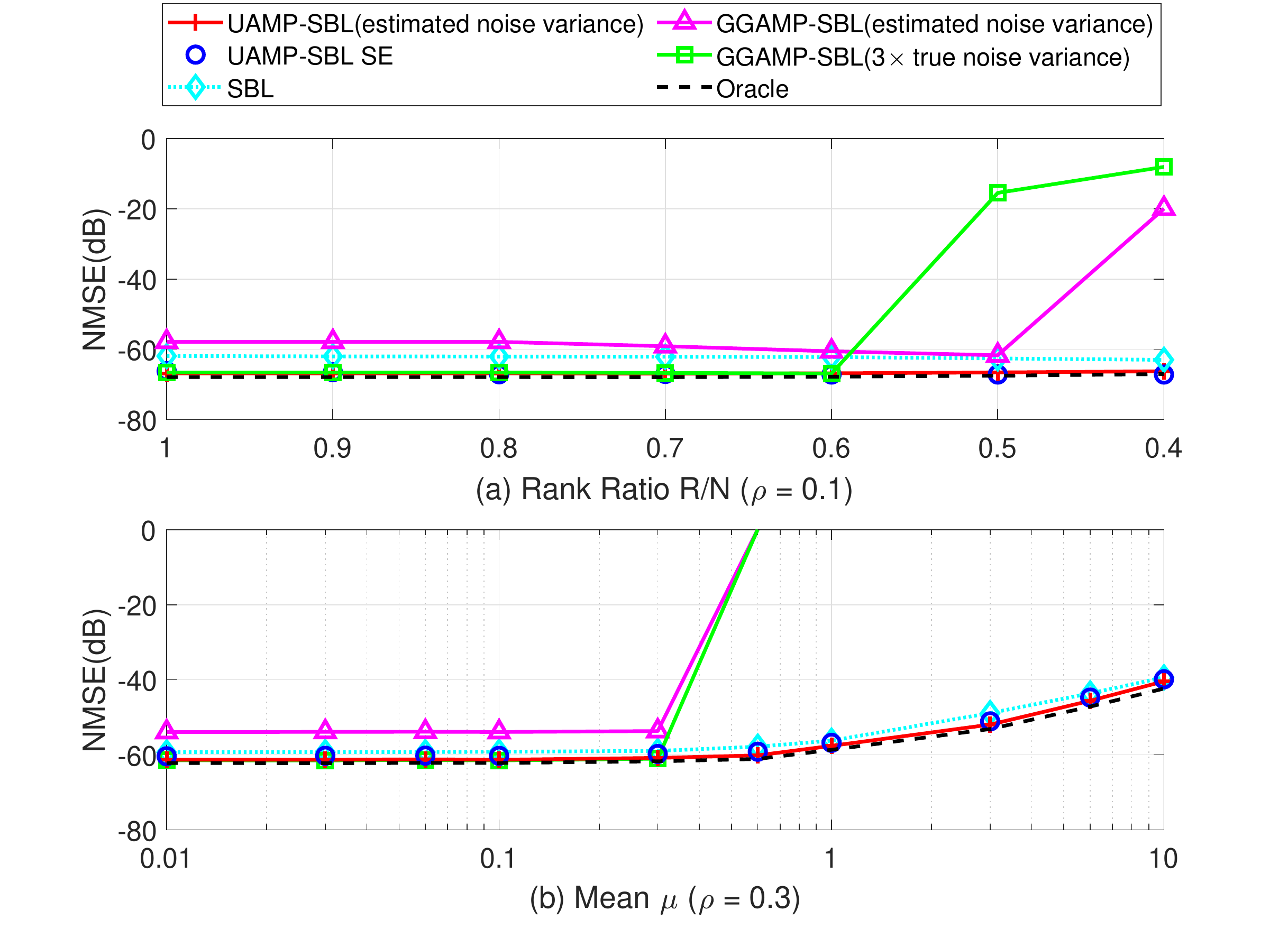}
	\caption{Performance comparison: (a)low rank matrices; (b) non-zero mean matrices.}
	\label{nonzeroresult}
\end{figure}



In Fig.~\ref{conditionresult}, the performance of various algorithms in terms of NMSE versus the condition number is shown in (a) for a sparsity rate of $\rho=0.1$ and (b) for a sparsity rate of $\rho=0.3$.  It can be seen from Fig.~\ref{conditionresult}(a) that UAMP-SBL delivers the best performance (even better than the conventional SBL algorithm), which closely approaches the support-oracle bound. With a larger sparsity rate in Fig.~\ref{conditionresult}(b), UAMP-SBL still exhibits excellent performance and it performs slightly better than SBL and significantly better than GGAMP-SBL when the condition number is relatively large. In addition, the simulation performance of UAMP-SBL matches well with the performance predicted with SE. 

Fig.~\ref{correlationresult} shows the performance of various algorithms versus a range of correlation parameter $c$ from $0.1$ to $0.5$, where the sparsity rate $\rho=0.1$ in (a) and $\rho=0.3$ in (b). \rea{It can be seen that}, UAMP-SBL still delivers exceptional performance, which is better than SBL and significantly better than GGAMP-SBL when the correlation parameter $c$ is relatively large. The gap between UAMP-SBL and GGAMP-SBL becomes more notable with a higher sparsity rate. The performance of UAMP-SBL matches well with SE again.

In Fig.~\ref{nonzeroresult}, we examine the performance of the algorithms versus rank ratio in (a), where the sparsity rate $\rho=0.1$, and versus non-zero mean in (b), where the sparsity rate $\rho=0.3$. It can be seen that UAMP-SBL still delivers performance which closely matches the support-oracle bound, and is slightly better than that of SBL. We can also see that GGAMP-SBL diverges when the mean $\mu$ is relatively large. The performance of UAMP-SBL matches well with SE as well. 

\begin{figure}[t]
	\centering
	\includegraphics[width=0.95\linewidth]{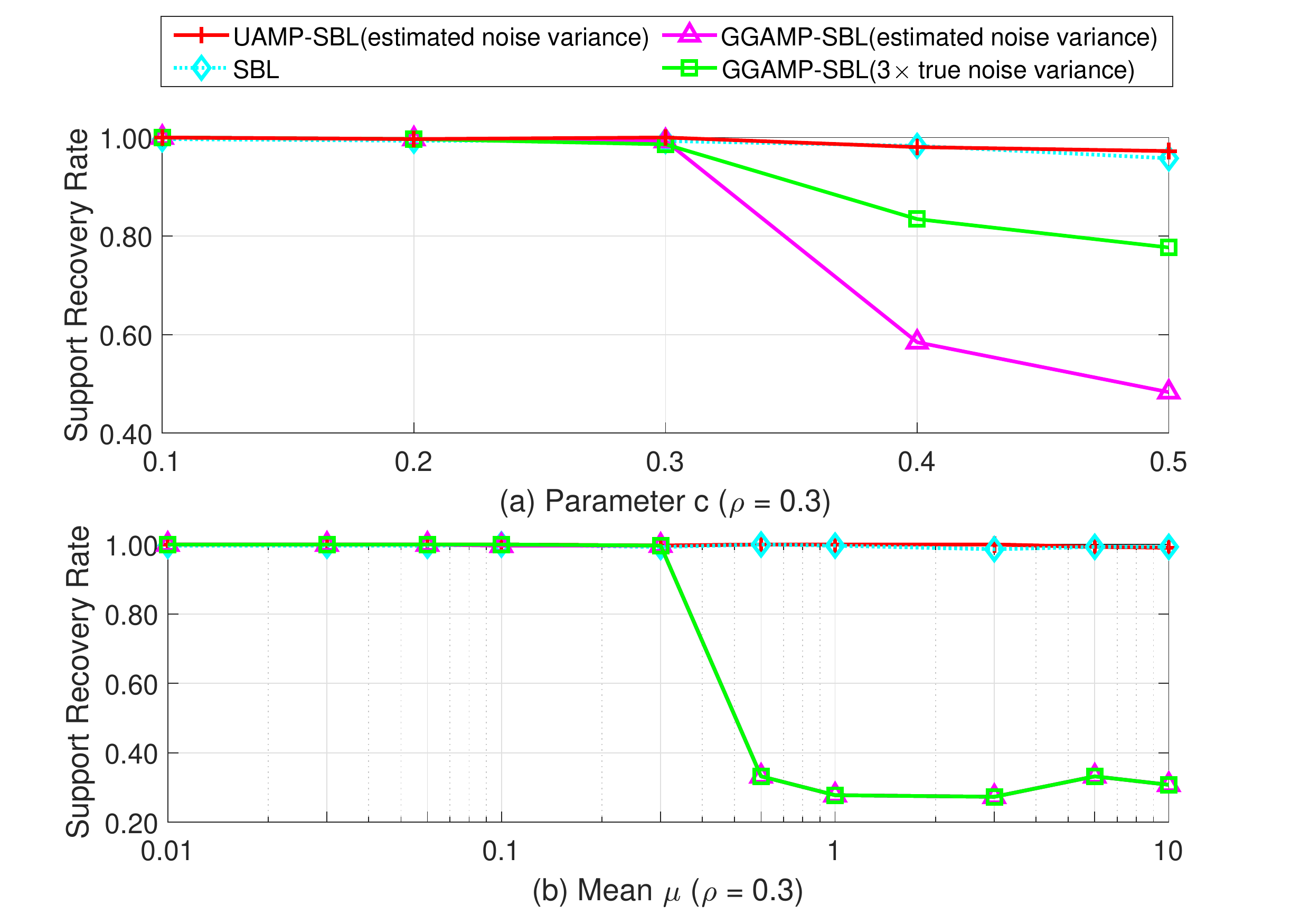}
	\caption{\rev{Support recovery rate comparison: (a) low rank matrices; (b) non-zero mean matrices.} }
	\label{supportrecoveryrate}
\end{figure}

\rev{In Fig. \ref{supportrecoveryrate}, we evaluate the support recovery rate of the algorithms versus correlation parameter $c$ for correlation matrices in (a) and mean value $\mu$ for non-zero mean matrices in (b), where the sparse rate $\rho=0.3$.  The support recovery rate is defined as the percentage of successful trials in the total trials \cite{sparserate}. In the noiseless case, a successful trial is recorded if the indexes of estimated non-zero signal elements are the same as the true indexes. In the noisy case, as the true sparse vector cannot be recovered exactly, the recovery is regarded to be successful if the indexes of the estimated elements with the $\mathcal{K}$ largest absolute values are the same as the true indexes of non-zero elements in the sparse vector $\mathbf{x}$, where $\mathcal{K}$ is the number of non-zero elements in $\mathbf{x}$. From the results, we can see that UAMP-SBL and SBL deliver similar performance and they can significantly outperform GGAMP-SBL when $c$ or $\mu$ is relatively large.}     

\begin{figure}[t]
	\centering
	\includegraphics[width=1.0\linewidth]{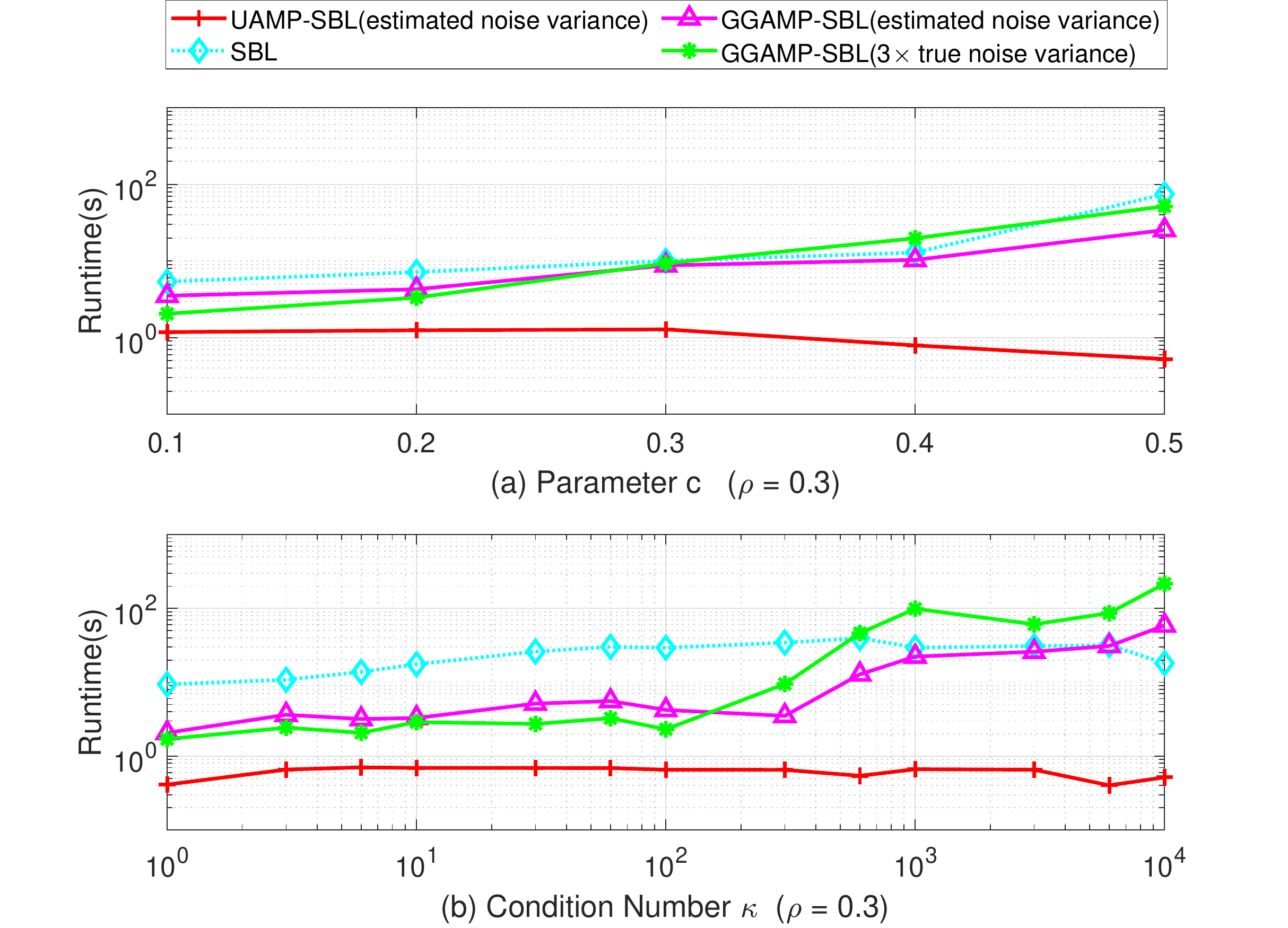}
	\caption{Runtime comparison: (a) correlated matrix; (b) ill-conditioned matrix.}
	\label{runtimerankresulta}
\end{figure}

The average runtime of various algorithms is shown in Figs.~ \ref{runtimerankresulta}, where the sparsity rate $\rho=0.3$, and the measurement matrice are correlated in (a) and ill-conditioned in (b). It can be seen that UAMP-SBL is much faster than GGAMP-SBL and SBL. SBL is normally the slowest as it has the highest complexity due to the matrix inverse in each iteration. It is noted that, for GGAMP-SBL, we set the damping factor to be relatively small value 0.2 to enable it to achieve better performance and robustness. If the damping factor is increased, GGAMP-SBL could become faster but at the cost of offsetting its performance and robustness.   

\begin{figure}[t]
	\centering
	\includegraphics[width=1.0\linewidth]{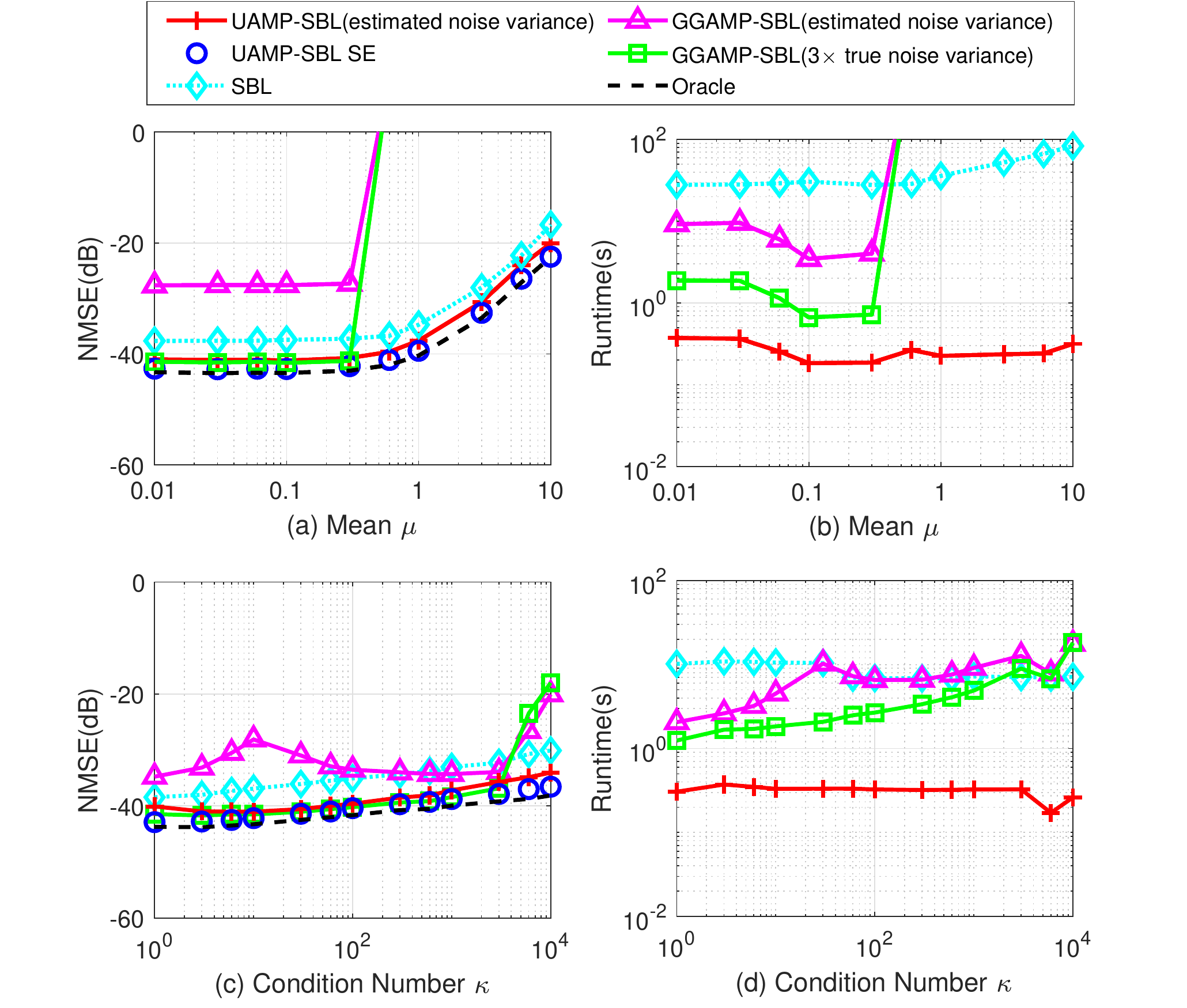}
	\caption{\rev{Performance and runtime comparisons of various algorithms where SNR = 35dB.} }
	\label{35dB}
\end{figure}

\rev{We also compare the performance of various algorithms at SNR = 35dB, and the NMSE performance and runtime of the algorithms are shown in Fig. \ref{35dB}, where (a) and (b) are for non-zero mean matrices, and (c) and (d) are for ill-conditioned matrices. The sparsity rate $\rho=0.1$.   Again, we can see that, compared to GGAMP-SBL, UAMP-SBL delivers better performance with considerably much smaller runtime when the mean or condition number of the matrices are relatively large.}   

\begin{figure}[t]
	\centering
	\includegraphics[width=1.0\linewidth]{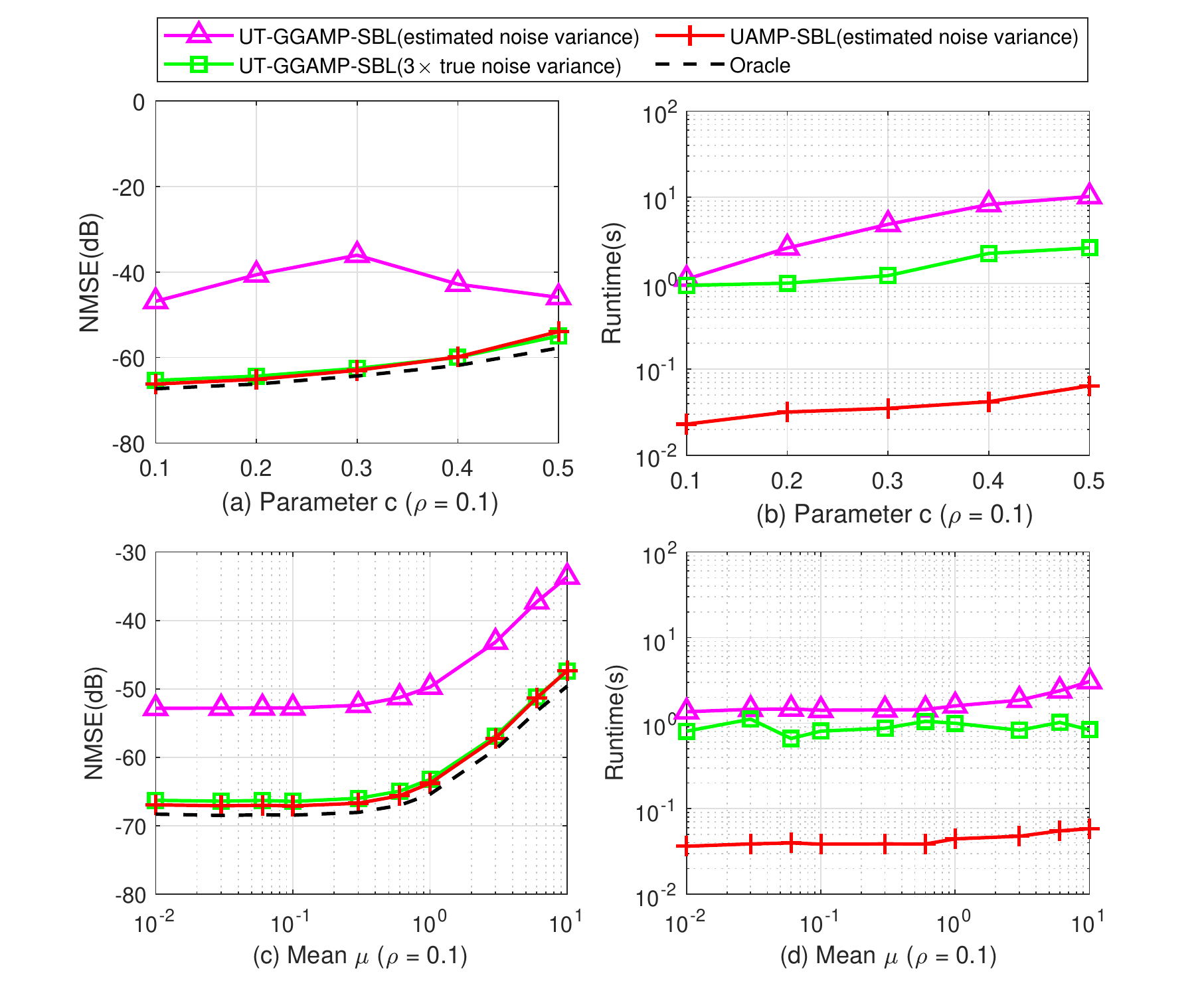}
	\caption{\rev{Performance and runtime comparisons of UAMP-SBL and UT-GGAMP-SBL.} }
	\label{utggampsbl}
\end{figure}

\rea{The key difference between AMP and UAMP is that a unitary transformation is performed in UAMP, which makes UAMP much more robust against a generic measurement matrix. Inspired by this, we test the impact of the unitary transformation on the GGAMP-SBL algorithm, where we first perform the unitary transformation to the original model and then carry out GGAMP-SBL. We call this algorithm UT-GGAMP-SBL, and compare it with UAMP-SBL. The performance and the corresponding runtime are shown in Fig. \ref{utggampsbl}, where (a) and (b) for correlated matrices, and (c) and (d) for non-zero mean matrices. It can be seen from this figure that, thanks to the unitary transformation, the stability of GGAMP-SBL is significantly improved as expected. UT-GGAMP-SBL with 3 times true noise variance achieves almost the same performance as UAMP-SBL, however, UT-GGAMP-SBL requires the knowledge of noise variance and it is significantly slower than UAMP-SBL.}

\begin{figure}[t]
	\centering
	\subfigure{
		\includegraphics[width=1.05 \linewidth]{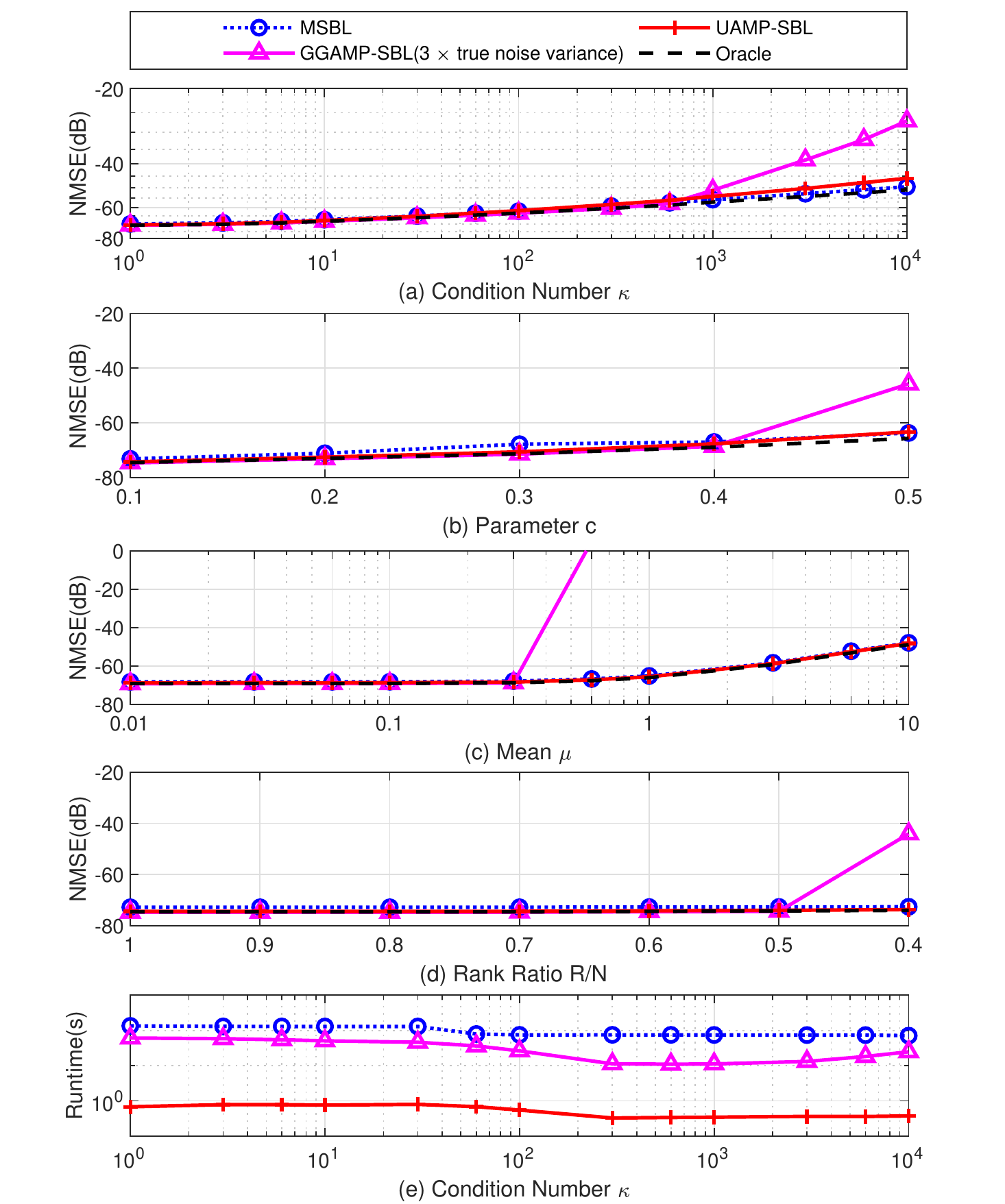}}
	\caption{Performance comparison of various algorithms \rev{in the case of MMV.}}
	\label{conditionresult_MMV}
\end{figure}

\subsection{Numerical Results for  MMV}

The elements of the sparse vectors $\left\lbrace  \mathbf{x}^{(l)},  l= 1:L\right\rbrace $ are drawn from a Bernoulli-Gaussian distribution, and the vectors share a common support. The number of measurement vectors is $5$. The performance of the algorithms with ill-conditioned, correlated, non-zero mean and low-rank measurement matrices is shown in Fig.~\ref{conditionresult_MMV} (a)-(d), respectively. In this figure, we also include the performance of {the direct extension of the conventional SBL algorithm to the MMV model (MSBL) \cite{wipf2007empirical} } and support-oracle bound.
It can be seen from this figure that, when the deviation of the measurement matrices from the i.i.d. zero-mean Gaussian matrix is small, GGAMP-SBL (with $3\times$ true noise variance) and UAMP-SBL deliver similar performance, and both of them can approach the bound closely. MSBL works slightly worse than GGAMP-SBL and UAMP-SBL. However, when the deviation is relatively large, MSBL delivers slightly better performance but at high complexity. In most cases, UAMP-SBL and MSBL almost have the same performance, and can significantly outperform GGAMP-SBL. As an example, we show the average runtime of different algorithms in the case of ill-conditioned matrices in Fig.~\ref{conditionresult_MMV}(e), \rev{where UAMP-SBL converges significantly faster than GGAMP-SBL and MSBL.}

%



 \begin{figure}[t]
	\centering
	\includegraphics[width=1.1\linewidth]{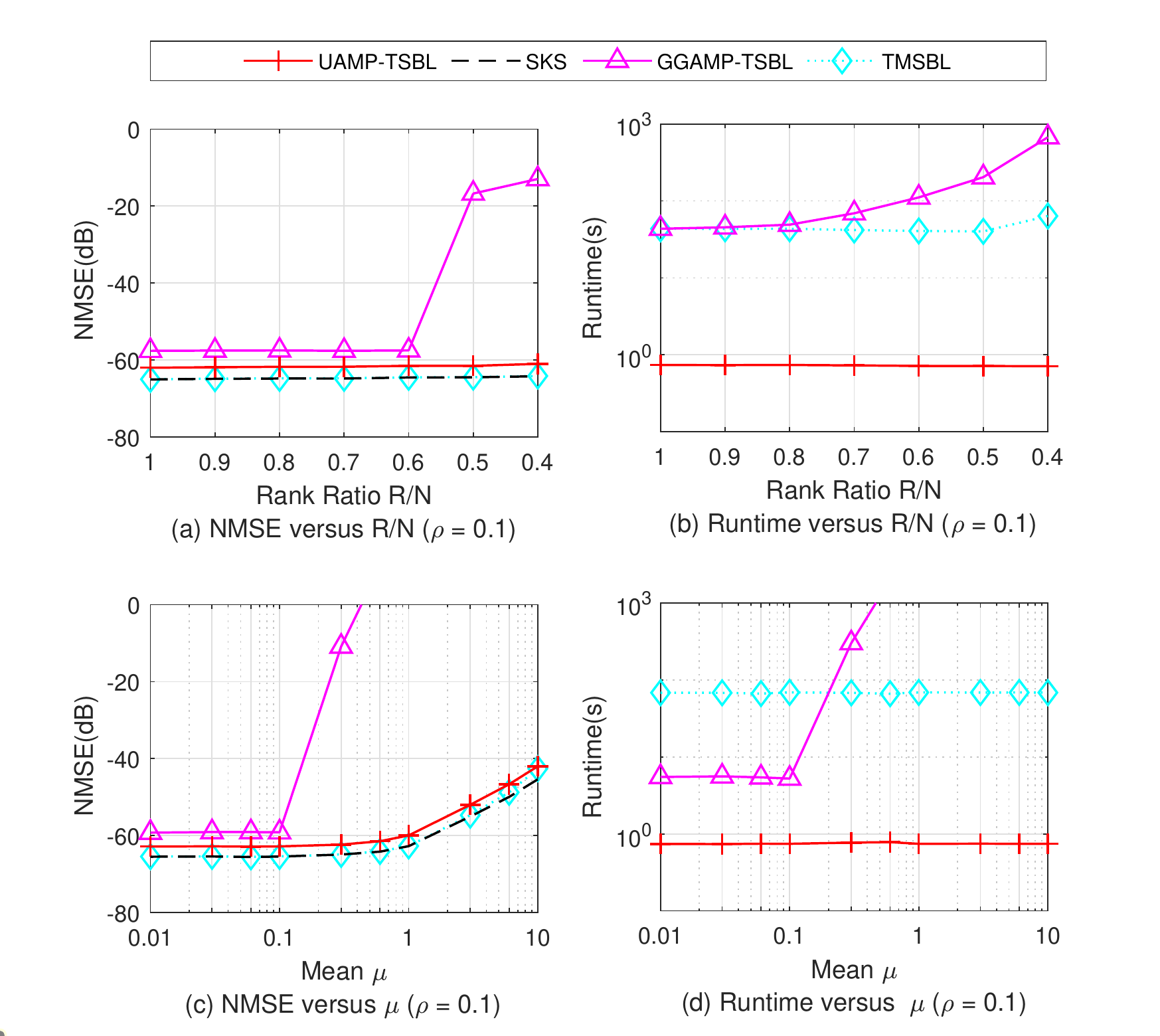} 
 	\caption{Performance comparison of various algorithms \rev{in the case of MMV with temporal correlation.}}
	\label{result_TMMV}
\end{figure}

Furthermore, we present a numerical study to illustrate the performance of UAMP-SBL when incorporating the temporal correlation. \rev{Besides the temporally correlated SBL (TMSBL) {\cite{sparserate}} and GGAMP-SBL, we also} compare the recovery performance with a lower bound: the achievable NMSE by a support-aware Kalman smoother (SKS) \cite{eubank2005kalman} with the knowledge of the support of the sparse vectors and the true values of $\beta$, $\alpha$ and $\bm{\gamma}$. 
The SKS is implemented in a more efficient way by incorporating UAMP. 
As examples, we use low rank and non-zero mean measurement matrices to test their performance. \rev{ The sparsity rate $\rho=0.1$, SNR = 50dB and the temporal correlation coefficient \rev{$\alpha = 0.8$}.} It can be seen from Fig.~\ref{result_TMMV} that, UAMP-TSBL can approach the bound closely and outperform other algorithms significantly when the rank ratio is relatively low and the mean is relatively high. In addition, UAMP-TSBL is much faster.

\section{Conclusion}

In this paper, leveraging UAMP, we proposed UAMP-SBL for sparse signal recovery with the framework of structured variational inference, which inherits the low complexity and robustness of UAMP against \rev{a generic measurement} matrix. We demonstrated that, compared to the state-of-the-art AMP based SBL algorithm, UAMP-SBL can achieve much better performance in terms of robustness, speed and recovery accuracy. \rev{Future work includes rigorous analyses of the state evolution of UAMP-SBL and the update mechanism of the shape parameter.}

\begin{appendices}
\section{Derivation of UAMP-SBL with SVMP}

We detail the forward and backward \rev{message passing in each subgraph of the factor graph in Fig. \ref{fig:factor graph}} according to the principle of SVMP \cite{jordan1999introduction}, \cite{winn2005variational},  \cite{dauwels2007variational}. \rev{The notation $\mathcal{M}_{n_a \rightarrow  n_b} (x)$ is used to denote a message passed from node $n_a$ to node $n_b$,} which is a function of $x$. \rev{Note that, if a forward message computation requires backward messages, we use the
messages in previous iteration by default.}

\subsubsection{Message Computations in Subgraph 1}  In this subgraph, we only need to compute the outgoing (forward) messages $\{\mathcal{M}_{\beta \rightarrow  f_{r_m}} (\beta)\}$, which are input to Subgraph 2. The derivation of the message update rule is delayed in the message computations in Subgraph 2, and is given in \eqref{subgraph1c}.

\subsubsection{Message Computations in Subgraph 2} According to SVMP, we need to run BP in this subgraph except at the factor nodes $\{f_{r_m}\}$ as they connect external variable nodes. Due to the involvement of $\mathbf{\Phi}$, this is the most computational intensive part, \rea{and we propose to use UAMP to handle it by integrating it to the message passing process.} 

According to the derivation of (U)AMP using loopy BP, UAMP provides the message \rea{from variable node $h_m$ to function node  $f_{r_m}$. Due to the  Gaussian approximation in the the derivation of (U)AMP, the message is Gaussian, i.e.,}
\begin{equation}
\mathcal{M}_{{h_m}\rightarrow f_{r_m}} (h_m) = \mathcal{M}_{f_{\delta_m}\rightarrow h_m} (h_m) = \mathcal{N} (h_m |  p_m, \tau_{p_m}),
\end{equation} 
where the mean $p_m$ and the variance $\tau_{p_m}$ \rea{are respectively the $m$th elements of $\bm{p}$ and $\bm{\tau}_{p}$  given in} Line 2 and Line 1 of the UAMP algorithm (Algorithm \ref{UTAMPv1}), 
which are also Line 2 and Line 1 of the UAMP-SBL algorithm (Algorithm \ref{vector_UTAMP_SBL_SMV-table}).

Following SVMP {\cite{dauwels2007variational}}, the message $\mathcal{M}_{f_{r_m}\rightarrow \beta} (\beta) $ from factor node $f_{r_m}$ to variable node ${\beta}$ can be expressed as
\begin{equation}
\mathcal{M}_{f_{r_m}\rightarrow \beta} (\beta) \propto \exp \left\{  \left\langle {\log  f_{r_m} ( {{r_m| h_m, \beta^{-1}}}  )} \right\rangle_{b\left( {h_m} \right) } \right\},\\
\label{addnewa}
\end{equation}
where \rea{the belief of $h_m$ is given as  
\begin{equation}	
b(h_m) \propto \mathcal{M}_{{h_m}\rightarrow f_{r_m}} (h_m) \mathcal{M}_{f_{r_m}\rightarrow h_m} (h_m).  
\end{equation}}
Later we will see that $
\mathcal{M}_{f_{r_m}\rightarrow h_m} (h_m) \propto \mathcal{N} (h_m | r_m, \hat{ \beta}^{-1})$ where $\hat{ \beta}^{-1}$ is an estimate of $\beta^{-1}$ (in the last iteration), and its computation is delayed to (\ref{n1}). Hence  $b(h_m )$ is Gaussian \rea{according to the property of the product of Gaussian functions}, i.e., $b(h_m)  = \mathcal{N} (h_m | \hat{h}_m, v_{h_m})$ with
\begin{eqnarray}
v_{h_m} &=& ({ {1}/{\tau_{p_m}} +\hat{ \beta} })^{-1} \\
\hat{h}_m &=&  v_{h_m} ( \hat{ \beta} {r_m}   +{p_m}/{\tau_{p_m}}).
\end{eqnarray} 
They can be rewritten in vector form as
\begin{eqnarray}
\mathbf{v}_h&=&\bm{\tau}_p./(\bm{1}+\hat\beta\bm{\tau}_p) \\
\mathbf{\hat h}&=&(\hat\beta\bm{\tau}_p\cdot \bm{r}+\bm{p})./(\bm{1}+\hat\beta\bm{\tau}_p),
\end{eqnarray}
to avoid numerical problems as $\bm{\tau}_p$ may contain zero elements, which are Lines 3 and  4 of the UAMP-SBL algorithm. Then, from \eqref{addnewa} \rea{and the Gaussianity of $b(h_m)$}, the message $\mathcal{M}_{f_{r_m}\rightarrow \beta} (\beta) $ is
\begin{equation}
\mathcal{M}_{f_{r_m}\rightarrow \beta} (\beta) \propto \sqrt{\beta} \exp \left\{   { -\frac{\beta}{2} } { ( |  r_m - \hat{h}_m   |^2 + v_{h_m})} \right\}. 
\end{equation}
According to SVMP, \rea{the message from function node $f_{r_m}$ to variable node $h_m$ is}
\begin{equation}
\begin{aligned}
\mathcal{M}_{f_{r_m}\rightarrow h_m} (h_m)
& \propto \exp \left\{  \left\langle {\log  f_{r_m} ( {{r_m| h_m, \beta^{-1}}}  )} \right\rangle_{b\left( {\beta} \right) } \right\}\\
&\propto \mathcal{N} (h_m |  r_m, \hat{ \beta}^{-1})
\label{M_frm_to_hm},
\end{aligned} 
\end{equation}
where $\hat{ \beta}= \left\langle \beta \right\rangle _{b(\beta)}$ with
\begin{eqnarray} 
b(\beta)&\!\!\!\!\!=\!\!\!\!\!& \mathcal{M}_{\beta\rightarrow f_{r_m}} (\beta) \mathcal{M}_{f_{r_m}\rightarrow \beta}(\beta) \nonumber \\ 
&\!\!\!\!\!=\!\!\!\!\!& f_{\beta} (\beta)  \prod_{m} \mathcal{M}_{f_{r_m}\rightarrow \beta} (\beta) \nonumber \\
&\!\!\!\!\!\propto\!\!\!\!\!&  {\beta}^{\frac{M}{2}-1} \exp \left\{   { -\frac{\beta}{2} } \sum_{m} { \left( | r_m - \hat{h}_m |^2 + v_{h_m}  \right) } \right\}, 
\end{eqnarray}
and
\begin{eqnarray} 
\mathcal{M}_{\beta\rightarrow f_{r_m}} (\beta) = f_{\beta} (\beta)  \prod_{m'\neq m} \mathcal{M}_{f_{r_{m'}}\rightarrow \beta} (\beta). 
\label{subgraph1c}
\end{eqnarray}

\rea{It is noted that $b(\beta)$ is a Gamma distribution with the rate parameter $ \frac{1}{2}\sum_m \left(|r_m-\hat{h}_m |^2+v_{h_m}\right) $ and the shape parameter $ M/2$, so $\hat{ \beta}= \left\langle \beta \right\rangle _{b(\beta)}$ can be computed as }  
\begin{equation}
\hat{ \beta}  =  {M}/{\sum_{m} { \left( | r_m - \hat{h}_m  |^2 + v_{h_m}  \right) }},\label{n1}
\end{equation} 
which can be rewritten in vector form shown in Line 5 of the UAMP-SBL algorithm. 

From \eqref{M_frm_to_hm}, the Gaussian form of the message $\mathcal{M}_{f_{r_m}\rightarrow h_m} (h_m)$ suggests the following model 
\begin{equation}
r_m=h_m+w_m, m=1, ..., M,
\end{equation} 
where $w_m$ is a Gaussian noise with mean 0 and variance $\hat \beta^{-1}$. This fits into the forward recursion of the UAMP algorithm as if the noise variance is known. Therefore, Lines 3 - 6 of the UAMP algorithm (Algorithm \ref{UTAMPv1}) can be executed, which are Lines 6 - 9 of the UAMP-SBL algorithm. According to the derivation of (U)AMP, UAMP produces the message { $\mathcal{M}_{{x_n}\rightarrow f_{x_n}} (x_n) \propto \mathcal{N} (x_n |  q_n, \tau_q)$} with mean $q_n$ and variance $\tau_q$, which are given in Lines 5 and 6 of the UAMP algorithm or Line 8 and Line 9 of the UAMP-SBL algorithm. \rea{We can see that the UAMP algorithm is integrated.}

The function nodes $\left\lbrace  f_{x_n}\right\rbrace $ connect the external variable node $\gamma_n$. According to SVMP, the outgoing message of Subgraph 2 $\mathcal{M}_{f_{x_n}\rightarrow {\gamma_n}} (\gamma_n)$ can be expressed as
\begin{eqnarray}
\mathcal{M}_{f_{x_n}\rightarrow {\gamma_n}} (\gamma_n) 
\propto   \exp \left\{  \left\langle  { {\log f_{x_n} ( {{x_n| 0, \gamma_n^{-1}}}  )}   } \right\rangle _{b\left( {x_n} \right)   } \right\}, 
\label{M_fx_ga}
\end{eqnarray}
where \rea{the belief $b(x_n) \propto \mathcal{M}_{{x_n}\rightarrow f_{x_n}} (x_n)  \mathcal{M}_{f_{x_n}\rightarrow x_n} (x_n)$.}

The message $\mathcal{M}_{f_{x_n}\rightarrow x_n} (x_n) \propto  \mathcal{N}  ( {{x_n| 0, \hat{ \gamma}_n^{-1}}}  )$  will be computed in (\ref{M_fx_x}),  where $\hat{ \gamma}_n= \left\langle \gamma_n \right\rangle _{b(\gamma_n)}$. Then \rea{$b(x_n)$ turns out to be Gaussian, i.e.,} $b(x_n)=\mathcal{N} (  {x_n}  | \hat{x}_n, \tau_{x_n})$ with
\begin{eqnarray}
\tau_{x_n}  = \left( {1}/{\tau_q} +\hat{ \gamma}_n \right) ^{-1} 
\label{xn} \\
\hat{ x }_n =  {q_n} /{(1+ \tau_q \hat{ \gamma}_n )}. 
\label{taux}
\end{eqnarray}
Performing the average operations to $\{\tau_{x_n}\}$ in \eqref{xn} and arranging \eqref{taux} in a vector form lead to Lines 10 and 11 of the UAMP-SBL algorithm. According to the above, 
\begin{eqnarray}
\mathcal{M}_{f_{x_n}\rightarrow {\gamma_n}} (\gamma_n) \propto \sqrt{\gamma_n} \exp \left\{  -\frac{\gamma_n}{2} {({  |\hat{x}_n|^2 + \tau_x} )   } \right\},
\label{M_fx_ga}
\end{eqnarray}
which is passed to Subgraph 3. 
This is the end of the message update in Subgraph 2.  

\subsubsection{Message Computations in Subgraph 3}   The message  $\mathcal{M}_{f_{\gamma_n} \rightarrow {\gamma_n}} (\gamma_n)$ from the factor node  $f_{\gamma_n}$  to the variable node $\gamma_n$ is a predefined Gamma distribution with shape parameter $\epsilon$ and rate parameter $\eta$, i.e.,   
\begin{equation}
\mathcal{M}_{f_{\gamma_n}\rightarrow \gamma_n} (\gamma_n)
\propto {\gamma_n}^{{\epsilon}-1}  \exp \left\{-{\eta}{\gamma_n}\right\}.
\label{M_fgn_to_gn}
\end{equation}
According to SVMP, \rea{the message } 
\begin{equation}
\mathcal{M}_{f_{x_n}\rightarrow x_n} (x_n) 
\propto  \exp \left\{  \left\langle  { {\log  f_x ( {{x_n| 0, \gamma_n^{-1}}}  )}   } \right\rangle _{b\left( {\gamma_n} \right)   } \right\} , 
\label{M_fx_x}
\end{equation}   
where \rea{the belief of $\gamma_n$}
\begin{equation}
\begin{aligned}
b(\gamma_n)
&\propto   \mathcal{M}_{f_{\gamma_n}\rightarrow \gamma_n} (\gamma_n)  \mathcal{M}_{f_{x_n}\rightarrow \gamma_n} (\gamma_n)  \\
&\propto    {\gamma_n}^{{\epsilon} - \frac{1}{2}}  \exp \left\{  -\frac{\gamma_n}{2} {  ( {|\hat{x}_n|^2 + \tau_x  +2\eta} )   } \right\}.
\end{aligned}
\end{equation}
Hence, the message 
\begin{equation}
\mathcal{M}_{f_{x_n}\rightarrow x_n} (x_n) 
\propto \mathcal{N} (  {x_n}  |{ 0 },\hat{\gamma}_n^{-1} ) ,
\label{M_fx_x}
\end{equation}
where 
\begin{equation}
\hat{\gamma}_n =\left\langle  {  {\gamma_n}} \right\rangle_{b ( {\gamma_n}) } 
= \frac{ {2 {\epsilon} +1}}{ { 2\eta +{{|\hat{x}_n|^2 + \tau_x}  }} } .
\label{ga_smv}  
\end{equation}
Here we set $\eta=0$, and $\hat{\gamma}_n$ is reduced to 
$\frac{(2{\epsilon}+1)}{{|\hat{x}_n|^2 + \tau_x}}$, which leads to Line 12 of the UAMP-SBL algorithm.  

We propose to tune the parameter automatically with the  empirical update rule for $\epsilon$ shown in Line 13 of the UAMP-SBL algorithm. The iteration is terminated when either the difference between two consecutive estimates of $\mathbf{x}$ is smaller than a threshold or the iteration number reaches the pre-set maximum value $t_{max}$.

\section{Proof of Proposition 1}        
    When  $\epsilon = 0$, 
	the iteration in terms of $\gamma_n$ has a simplified closed form,
	i.e.,   
	\begin{align}
	\gamma^{t+1}_n = g_{\epsilon_0}(\gamma^t_n)
	=  \frac{ (\beta + \gamma^t_n  )^{2} }{(\beta y_n )^2 +\beta + \gamma^t_n }.  
	\label{iteration_epsilon=0}
	\end{align}
	In order to find the fixed point, we need to solve the following equation
	\begin{equation}
	f(\gamma_n)={ g_{\epsilon_0}(\gamma_n) }  - \gamma_n = 0,
	\end{equation} 
	\rev{which leads to the unique root}
	\begin{align}
	\gamma'_{n} =  \frac{\beta}{\beta y^2_n   - 1 }.
	\end{align}
	
	If $\beta y^2_n > 1$, the root $\gamma'_{n} = \frac{\beta}{\beta y^2_n   - 1 } >0$.
	Taking the derivative of $g_{\epsilon_0}(\gamma_n)$ in (\ref{iteration_epsilon=0}), we have 
	\begin{align}
	\frac{{d} }{{d} \gamma_n  } g_{\epsilon_0}(\gamma_n)  
	&=  1- \left(\frac{    {\beta}^2 y^2_n   }{ {\beta}^2 y^2_n  +\beta + \gamma_{n} }\right )^2.
	\label{g'(gamma_1)}
	\end{align}
	\rev{It is easy to verify that, when $\gamma_n >0$, $0<\frac{{d} }{{d} \gamma_n  } g_{\epsilon_0}(\gamma_n) <1$. Thus, the unique root $\gamma'_{n} =  \frac{\beta}{\beta y^2_n - 1}$ is a stable fixed point of the iteration. As $0<\frac{{d} }{{d} \gamma_n  } g_{\epsilon_0}(\gamma_n) <1$ when $\gamma_n>0$, with an initial value $\gamma^{(0)}_n>0$, $\gamma^{t}_n$ will converge to the stable fixed point  $\gamma'_{n}$ \cite{book1}.}     
	
	\rev{If $\beta y^2_n \leq 1$, the root $\gamma'_{n} = \frac{\beta}{\beta y^2_n - 1} <0$ or $\gamma'_{n} = +\infty$, i.e., there is no cross-point between $y=g_{\epsilon_0}(\gamma_n)$ and $y =\gamma_n$ when $\gamma^n >0$. As $g_{\epsilon_0}(0) = \frac{ \beta^{2} }{(\beta y_n)^2 +\beta } > 0$, $y=g_{\epsilon_0}(\gamma_n)$ is above $y =\gamma_n$ for $\gamma_n>0$. In addition, $y=g_{\epsilon_0}(\gamma_n)$ is an increasing function for $\gamma_n>0$. Hence $\gamma_n^t$ goes to $+\infty$ with the iteration.}         

\section{Proof of Theorem 1}

With $\epsilon>0$, the derivative of $g_{\epsilon}(\gamma_n)$ is given as 
\begin{align}
\frac{d g_{\epsilon}(\gamma_n) }{d \gamma_n  } 
= (2\epsilon +1)   \left( 1- \left(\frac{   \beta  u_n   }{ \beta  u_n +\beta + \gamma_{n}  }\right )^2  \right)  
\label{g'(ga)},
\end{align}
where $ u_n = \beta y_n^2 $. 
\rea{To find the fixed points of the iteration, we let} $f(\gamma_n)={ g_{\epsilon}(\gamma_n) }  - \gamma_n = 0$,
\rea{leading to}
\begin{equation}
2\epsilon \gamma^2_n - \gamma_n \beta(\beta y^2_n -4\epsilon -1) +\beta^2(1+2\epsilon)=  0. 
\label{mother_of_gamma_roots}
\end{equation}
The two roots of (\ref{mother_of_gamma_roots}) are given by \footnote{\rea{An alternative form for the quadratic formula is used, which can be deduced from the standard quadratic formula by Vieta's formulas.}}
\begin{equation}
\gamma_{n(a)} =  \frac{ 2\beta(1+2\epsilon)}{u_n -4\epsilon - 1 + \sqrt{ u_n^2 - 8\epsilon u_n - 2u_n +1  } },
\label{ga1}
\end{equation} 
and
\begin{equation}
\gamma_{n(b)} = \frac{ 2\beta(1+2\epsilon)}{u_n -4\epsilon - 1 - \sqrt{ u_n^2 - 8\epsilon u_n - 2u_n +1  } }.
\label{ga2}
\end{equation}
If $u_n > 1+ 4\epsilon +4\sqrt{\epsilon^2 + \epsilon/2} $, it is not hard to verify that $u_n -4\epsilon - 1 - \sqrt{ u_n^2 - 8\epsilon u_n - 2u_n +1}>0$, so both roots are positive. Hence they are two fixed points of the iteration. Next, we show that $\gamma_{n(a)}$ is a stable fixed point while $\gamma_{n(b)}$ is an unstable one.   

Plugging the root $\gamma_{n(a)}$ into  \eqref{g'(ga)}, we have
\begin{align}
\frac{{d} }{{d} \gamma_n  } g_{\epsilon}(\gamma_n) \Bigg|_{\gamma_n = \gamma_{n(a)}} 
&=(2\epsilon +1)   \left( 1- \left(\frac{   \beta  u_n   }{ \beta  u_n +\beta + \gamma_{n(a)}  }\right )^2  \right) .
\label{g'(gamma_1)}
\end{align}
It is clear that the derivative is larger than 0. Verifying that $\frac{{d} }{{d} \gamma_n} g_{\epsilon}(\gamma_n) |_{\gamma_n = \gamma_{n(a)}}<1$ is equivalent to showing that
\begin{equation}
l(u_n) =  (2\epsilon+1)(\beta u_n)^2 - 2\epsilon( \beta  u_n +\beta + \gamma_{n(a)} )^2 
\label{q(h)}
\end{equation}
is larger than 0.
Inserting \eqref{ga1} into \eqref{q(h)},
\begin{equation}
\begin{aligned}
\rev{\frac{4\epsilon l(u_n)}{\beta^2 }}
& = l_1(u_n)  +( (4\epsilon + 1) u_n  -1 ) \sqrt{ -l_1(u_n) },  
\end{aligned}
\end{equation}
where
\begin{equation}
l_1(u_n) = - (u_n^2 - 8\epsilon u_n - 2u_n +1 ) <0.
\end{equation}
Then
\begin{equation}
\begin{aligned}
\rev{\frac{4\epsilon l(u_n)}{\beta^2 }}\! =\!  \sqrt{-l_1(u_n)}  \big( - \sqrt{-l_1(u_n)} \!+\! (4\epsilon u_n \!+\! u_n \!-\!1) \big).
\end{aligned}
\label{stable}
\end{equation}
Because
\begin{equation}
\begin{aligned}
(4\epsilon u_n + u_n -1)^2- {(-\rev{l_1}(u_n))}=16\epsilon^2 u_n^2 +8\epsilon u_n >0,
\end{aligned}
\label{inequailty_for_mods}
\end{equation}
\rev{the term in \eqref{stable} $- \sqrt{-l_1(u_n)} + (4\epsilon u_n + u_n -1) >0 $ and we have $l(u_n)>0$.
Therefore, $\frac{{d} }{{d} \gamma_n} g_{\epsilon}(\gamma_n) |_{\gamma_n = \gamma_{n(a)}}<1$, i.e., $\gamma_{n(a)}$ is a stable fixed point. Similarly, it is not hard to show that $l(u_n) <0$ (i.e., $ \frac{{d} }{{d} \gamma_n  } g_{\epsilon}(\gamma_n) >1 $) for $\gamma_n=\gamma_{n(b)}$, i.e., $\gamma_{n(b)}$ is an unstable fixed point.}  

\rev{\rea{Then} we analyze the convergence behavior. As $\gamma_n >0$, the derivative (\ref{g'(ga)}) is an increasing function and it is positive.  In the above, it is already shown that $\frac{{d} }{{d} \gamma_n} g_{\epsilon}(\gamma_n) |_{\gamma_n = \gamma_{n(a)}}<1$. Therefore, for $\gamma_n \in [0,\gamma_{n(a)}]$, $  0< \frac{{d} }{{d} \gamma_n  } g_{\epsilon}(\gamma_n)  < 1.$ Thus, with an initial $\gamma^{(0)}_n$ with the range, $\gamma^{t}_n$ converges to the stable fixed point $\gamma_{n(a)}$ \cite{book1}.}

\rev{Next we consider $ u_n < 1+ 4\epsilon +4\sqrt{\epsilon^2 + \epsilon/2} $. For $u_n \in (1+ 4\epsilon -4\sqrt{\epsilon^2 + \epsilon/2}, 1+ 4\epsilon +4\sqrt{\epsilon^2 + \epsilon/2})$, it can be verified that $  u_n^2 - 8\epsilon u_n - 2u_n +1 < 0 $, leading to two complex roots $\gamma_{n(a)}$ and  $\gamma_{n(b)}$. 
If $ u_n \leq 1+ 4\epsilon -4\sqrt{\epsilon^2 + \epsilon/2} $, it can be shown that $  u_n^2 - 8\epsilon u_n - 2u_n +1 \geq 0 $ and $u_n^2 - 8\epsilon u_n - 2u_n +1 < (u_n - 4\epsilon- 1)^2$. Thus $u_n - 4\epsilon- 1 < -4\sqrt{\epsilon^2 + \epsilon/2} <0$ and $ u_n - 4\epsilon- 1 \pm \sqrt{u_n^2 - 8\epsilon u_n - 2u_n +1 }<0$, leading to negative $\gamma_{n(a)}$ and  $\gamma_{n(b)}$. In summary, 
if $u_n < 1+ 4\epsilon +4\sqrt{\epsilon^2 + \epsilon/2} $,  the two roots are either complex or negative. Hence, there is no cross-point between $y=g_{\epsilon}(\gamma_n)$ and $y =\gamma_n$ for  $\gamma_n>0$. As  $g_{\epsilon}(0) = (2\epsilon + 1) \frac{ \beta^{2} }{(\beta y_n)^2 +\beta } > 0$, $y=g_{\epsilon}(\gamma_n)$ is above $y=\gamma_n$. Meanwhile $g_{\epsilon}(\gamma^t_n)$ is an increasing function. Hence, $\gamma^t_n$ goes to $+\infty$ with the iteration.}    

\rev{When $u_n = 1+ 4\epsilon +4\sqrt{\epsilon^2 + \epsilon/2}$, there is single root $\gamma^*_n = \frac{2\beta(1+2\epsilon)}{u_n - 1- 4\epsilon}$. Plugging $\gamma^*_n$ into (\ref{g'(ga)}), we have$\frac{d g_{\epsilon}(\gamma_n) }{d \gamma_n} = 1$. Thus $\gamma^*_n$ is neutral fixed point \cite{book1}. Depending on the initial value $\gamma^{(0)}_n$, $\gamma^t_n $  may converge to the fixed point $\gamma^*_n$ or diverge.}

\section{Derivation of UAMP-SBL for MMV}

The \rea{belief} $b(\beta)$ can be represented as
\begin{eqnarray}
b(\beta)&\propto& f_{\beta} (\beta)  \prod_{l,m} { \mathcal{M}_{f^{(l)}_{r_m} \rightarrow  {\beta}}} (\beta)   \nonumber \\
&\propto&  1/\beta \prod_{l,m} \mathcal{N} ({h}^{(l)}_m |  {r}^{(l)}_m, \hat{ \beta}^{-1}).
\end{eqnarray}
\rea{Then according to $\hat\beta =<\beta>_{b(\beta)}$, we have}
\begin{eqnarray}
\hat \beta= {ML}/{\sum_{m,l} { \left( | r^{(l)}_m - \hat{h}^{(l)}_m   |^2 + v^{(l)}_{h_m}  \right) }}. 
\end{eqnarray}
According to the factor graph in  Fig.~\ref{fig:factor graph MMV}, the belief ${b\left( {{{\gamma}_n}} \right) }$ can be updated as
\begin{eqnarray}
&&\!\!\!\!\!\!\!b(\gamma^{(l)}_n)  \propto  \mathcal{M}_{f^{(l)}_{ \gamma_n \rightarrow \gamma^{(l)}_n }}  (\gamma^{(l)}_n)  
\mathcal{M}_{ f^{(l)}_{ {{x}_n}   \rightarrow  \gamma^{(l)}_n }} ( \gamma^{(l)}_n)   \nonumber \\
&&\!\!\!\!\!\!\! = (\gamma^{(l)}_n )^{{ \epsilon} -1+ \frac{1}{2} } \exp \left\{- \frac{\gamma^{(l)}_n}{2} (2\eta +   ({ |\hat{x}_n^{(l)} |^2 +{\tau}^{(l)}_x}) )\right\}. 
\end{eqnarray}
Here, we still set $\eta = 0$ and  the expectation of  $\gamma_n$ leads to 
\begin{eqnarray}
\hat\gamma_n = \frac{ 2 {\epsilon'} + 1 }{ (1/L)  \sum^L_{l=1} ({ |\hat{{ x}}_n^{(l)} |^2 +{\tau}^{(l)}_x})  }, 
\label{addnew}
\end{eqnarray}
where $\epsilon'=\epsilon/L$. By comparing \eqref{addnew} with \eqref{ga_smv}, the update of $\epsilon'$ can be expressed as
\begin{eqnarray}
{\epsilon'} =
\frac{1}{2}\sqrt{\log(\frac{1}{N}\sum_{n}{\hat{\gamma}_n})-\frac{1}{N}\sum_{n}{\log{\hat{\gamma}}_n}}.
\end{eqnarray}

\section{Derivation of UAMP-TSBL}

\rea{We only derive the message passing for the graph shown in Fig. \ref{fig:figtsbl}.} The message $\mathcal{M}_{f^{(l)}_{x_n}\rightarrow x^{(l)}_n} (x^{(l)}_n)$ is computed by the BP rule with the product of messages $ \{ \mathcal{M}_{f^{(l-1)}_{\delta_m}\rightarrow x^{(l-1)}_n} (x^{(l-1)}_n) ,\forall m \}$ defined in UAMP and  message $ \{ \mathcal{M}_{f^{(l-1)}_{\delta_m}\rightarrow x^{(l-1)}_n} (x^{(l-1)}_n) \}$, i.e.,
\begin{equation}
\begin{aligned}
&\mathcal{M}_{f^{(l)}_{x_n}\rightarrow x^{(l)}_n} (x^{(l)}_n)\\ 
&=  \left\langle  { f_{x^{(l)}_n}( x^{(l )})  } \right\rangle_{ \mathcal{M}_{f^{(l-1)}_{x_n}\rightarrow x^{(l-1)}_n}   \prod_{m} \mathcal{M}_{f^{(l-1)}_{\delta_m}\rightarrow x^{(l-1)}_n}  } 
\\
&\propto \mathcal{N} ( x^{(l)}_n|\xi^{(l)}_n, \psi^{(l)}_n),     
\end{aligned}
\end{equation}
which leads to Lines 1 to 6 of the UAMP-TSBL algorithm.
Similarly, the message $\mathcal{M}_{f^{(l+1)}_{x_n}\rightarrow x^{(l )}_n} (x^{(l )}_n)$ from factor node $f^{(l+1)}_{x_n}$ to variable node $ x^{(l )}_n$ is also updated by the BP rule 
\begin{equation}
\begin{aligned}
&\mathcal{M}_{f^{(l+1)}_{x_n}\rightarrow x^{(l)}_n} (x^{(l)}_n) \\ 
&=  \left\langle  { f_{x^{(l+1)}_n}(x^{(l+1))}} \right\rangle_{ \mathcal{M}_{f^{(l+2)}_{x_n}\rightarrow x^{(l+1)}_n}   \prod_{m} \mathcal{M}_{f^{(l+1)}_{\delta_m}\rightarrow x^{(l+1)}_n}   } 
\\
&\propto \mathcal{N} ( x^{(l)}_n|\theta^{(l)}_n, \phi^{(l)}_n),     
\end{aligned}
\end{equation}
leading to Lines 22 to 27 of the UAMP-TSBL algorithm.
We compute the \rea{belief} of variable $x^{(l)}_n$ by 
\begin{equation}
\begin{aligned}
&b(x^{(l)}_n) \propto 
\mathcal{M}_{f^{(l)}_{x_n}\rightarrow x^{(l)}_n}  
\mathcal{M}_{f^{(l+1)}_{x_n}\rightarrow x^{(l)}_n} 
\prod_{m} \mathcal{M}_{f^{(l )}_{\delta_m}\rightarrow x^{(l )}_n}  \\
&\propto \mathcal{N} ( x^{(l)}_n|\hat x^{(l)}_n, \tau_x^{(l)} ) 
\end{aligned}
\end{equation}
leading to Lines 19 to 20 of the UAMP-TSBL algorithm.
With the beliefs $ b ({ x^{(l)}_n}) $ and $ b ({ x^{(l-1)}_n}) $, the message $\mathcal{M}_{f^{(l)}_{x_n}\rightarrow \gamma_n} (\gamma_n)$ can be obtained as
\begin{eqnarray}
\begin{aligned}
\mathcal{M}_{f^{(l)}_{x_n}\rightarrow \gamma_n}( \gamma_n)
= \exp \left\{  \left\langle  { f^{(l)}_{x_n}(x^{(l)}_n|{\gamma_n}) } \right\rangle _{ b ({ x^{(l)}_n}) b ({ x^{(l-1)}_n})  } \right\}.  \\ 
\label{M_fx_ga_MMV}
\end{aligned}
\end{eqnarray}
Then, with the message $\mathcal{M}_{f_{\gamma_n}\rightarrow \gamma_n} (\gamma_n) $ in (\ref{M_fgn_to_gn}), \rea{the belief} $b(\gamma_n)
\propto   \mathcal{M}_{f_{\gamma_n}\rightarrow \gamma_n} (\gamma_n)  \mathcal{M}_{f_{x_n}\rightarrow \gamma_n} (\gamma_n)$.
\rea{Then the update of $\hat \gamma_n$} can be expressed as
\begin{equation}
\begin{aligned}
\hat{\gamma}_n = L(2\epsilon' + 1)/ ( 
|\hat{\mathbf{ x}}_n^{(1)} |^2 +{\tau}^{(1)}_x +
\frac{1}{\alpha^2} \sum^L_{l=2} ({ |\hat{\mathbf{ x}}_n^{(l)} |^2 +{\tau}^{(l)}_x})\\ +
\frac{\alpha^2}{1-\alpha^2} \sum^{L-1}_{l=1} ({ |\hat{\mathbf{ x}}_n^{(l)} |^2 +{\tau}^{(l)}_x}) -
\frac{2\alpha}{1-\alpha^2} \sum^{L }_{l=2} ({  \hat{\mathbf{ x}}_n^{(l)} \hat{\mathbf{ x}}_n^{(l-1)} }) 
).
\end{aligned}
\end{equation}

\end{appendices}

\bibliographystyle{IEEEtran}
\bibliography{IEEEabrv,mybib}

\end{document}